%% file: tcom.tex
\newif\ifSC
\SCtrue
\ifSC
\documentclass[onecolumn,draftclsnofoot,12pt]{IEEEtran}
\else
\documentclass[10pt,twocolumn]{IEEEtran}
\fi

\ifCLASSINFOpdf
\else
\usepackage[dvips]{graphicx}
\fi

\input{define.tex}

\begin{document}
	\title{On the Performance of the Primary and Secondary Links in a 3-D Underlay Cognitive Molecular Communication}
	
	\author{Nithin V. Sabu,  Neeraj Varshney
		and Abhishek K. Gupta
		\thanks{ N. V. Sabu and A. K. Gupta are with Indian Institute of Technology Kanpur, Kanpur UP 208016, India (Email:{ \{nithinvs,gkrabhi\}@iitk.ac.in}). 
			N. Varshney is with the Wireless Networks Division, National Institute of Standards and Technology, Gaithersburg, MD 20899 USA (Email: {neerajv@ieee.org}). 
		This research was supported by the Science and Engineering
		Research Board (India) under the grant SRG/2019/001459 and IITK under the grant IITK/2017/157.}}
	
	\maketitle
	
	\begin{abstract}
Molecular communication often involves coexisting links where certain links may have priority over others. In this work, we consider a system in three-dimensional (3-D) space with two coexisting communication links, each between a point transmitter and fully-absorbing spherical receiver (FAR), where the one link (termed primary) has priority over the second link (termed secondary). The system implements the underlay cognitive-communication strategy for the co-existence of both links, which use the same type of molecules for information transfer. Mutual influence of FARs existing in the same communication medium results in competition for capturing the information-carrying molecules. In this work, first, we derive an approximate hitting probability equation for a diffusion-limited molecular communication system with two spherical FARs of different sizes considering the effect of molecular degradation. The derived equation is then used for the performance analysis of primary and secondary links in a cognitive molecular communication scenario.  We show that the simple transmit control strategy at the secondary transmitter can improve the performance of the overall system. We study the influence of molecular degradation and decision threshold on the system performance. We also show that the systems parameters need to be carefully set to improve the performance of the system.  
	\end{abstract}
	
	\section{Introduction}
	Molecular communication (MC) is the communication between the transmitter and the receiver by using  \textit{information molecules} (IMs) as the carrier of information \cite{Suda2005b,Nakano2011}.
	Among different propagation mechanisms, molecular communication via diffusion (MCvD) systems is the most popular, mainly due to the ease of mathematical modeling and its energy efficiency. In MCvD systems, bio-nanomachines (nanomachines with biological components) can be used as the transmitter and receiver. Since the capabilities of individual bio-nanomachines may be limited to simple sensing and actuation, the internet of bio-nano things (IoBNT) \cite{Akyildiz2015} is envisioned to enable the interconnection of several bio-nanomachines to perform complex tasks. Applications of IoBNT include intra-body sensing and actuation, connectivity control, control of toxic gases and pollution in the environment \cite{Akyildiz2015}, and diagnosis and mitigation of infectious diseases \cite{Akyildiz2020}.	\par
	\textit{Related works: }
	In an MCvD system with multiple communication links, some communication links may have priority over others. For example, inside the human body, an artificial MC link would have to co-exist with existing biological links  that are inherent in the body, and these biological links have priority over the artificial link. Interference beyond a threshold from the MC link on any biological link may disrupt the human-body activities when both links use the same type of IMs. Similarly, a bio-nanomachine sensing the blood sugar level inside a diabetic person can have priority over the bio-nanomachine sensing the body temperature. In systems with multiple coexisting links with different priorities, the one with higher priority is termed primary link and the rest are termed secondary links. The goal here is to allow secondary link communication while ensuring a certain quality of serviec (QoS) at the primary. In wireless communication, these systems are called \textit{cognitive radio}, where information like channel conditions, messages, codebooks, etc., are utilized for the co-existence of the links\cite{Goldsmith2009,Gupta2016}. The three different strategies used in cognitive radio are \textit{underlay}, \textit{overlay}, and \textit{interweave}. In the underlay strategy, the secondary transmitter controls the transmit power to limit the co-channel interference at primary receiver \cite{Gupta2016}. In the overlay strategy, the secondary transmitter assists the primary communication via relaying or interference cancellation with the help of  transmit and channel information provided by the primary. In the interleaving strategy, the secondary transmitter transmits its information only when it does not sense any primary communication. \par
	
	Analogous paradigms have been recently proposed for molecular communication\cite{Alizadeh2017,Egan2018,Egan2019,estnew,reactivenew}. Authors in \cite{Alizadeh2017} studied an interweaving MC system where the secondary transmitter intelligently senses the absence of primary transmission in the primary transmitter communication. 
	A framework for a system with co-existing MC and the biochemical system was discussed in \cite{Egan2018} where some enzymes are introduced to reduce the inter-symbol interference (ISI) to the MC system, and the perturbation of the biochemical system was analyzed along with the equilibrium of the biochemical system. \cite{Egan2019} studied an underlay cognitive-communication consisting of biological systems and an MC link. However the work is limited to the time evolution of the concentration of molecules and scaling laws for the probability of a molecule to reach the receiver. The study in \cite{estnew} utilise Kullback-Leibler divergence between the distribution of molecules on the surface of biological system in the presence and absence of MC system for the coexistence constraint. The work in \cite{Egan2019,estnew} are initial models developed based on the theory of chemical reaction networks and many modeling complexities including receiver size, geometry, and the mutual influence between receivers are not considered. In \cite{reactivenew}, a passive receiver is used as the receiver and coexistence constraint is obtained with the help of reactive signaling.\par
	
	Unlike wireless communication, in MC with absorbing receivers, the links face interference from transmitters as well as receivers that are present in the same media. This is due to the fact that such receivers will cause  IMs corresponding to a communication link to get absorbed at itself, which will reduce the number of IMs reaching the desired receiver. Since multiple receivers coexist in cognitive systems, it is essential to characterize the impact of receivers on each other in order to evaluate the performance of these systems. Many previous works including \cite{Alizadeh2017,Egan2018,Egan2019,estnew} ignore the mutual influence of absorbing receivers due to lack of its analytic expressions in the literature. An approximate analytical expression for the hitting probability of an IM on any of the  FAR (centers distributed as Poisson point process) is given in \cite{Sabu2020b}. For an MCvD system with two FARs of equal size, an approximate equation for the hitting probability of an IM on each of the FAR was derived in our previous work \cite{Sabu2020a}. However, in cognitive systems, primary and secondary receivers can be of different sizes whose expressions are unknown in the past literature.  Therefore, it is essential to consider the difference in the size of receivers in the system model.\par
	In this paper, we try to bridge these gaps by first characterizing the hitting probability of IMs on each FAR and then provide an analytical framework to study an underlay cognitive MC system using the derived expressions.\par
	\textit{Contributions: }
	In this work, we develop an analytical framework to study a cognitive molecular communication system with a primary link and a secondary link implementing transmit control to limit interference at the primary link (\ie underlay). For the same, we first characterize the mutual impact of receivers on each other by deriving an approximate closed expression for the hitting probability of an IM on each of the FARs in the presence of the other FAR when they have different radii.  In this paper, we also consider that the IMs in the propagation medium can undergo molecular degradation \cite{Heren2015} in deriving the hitting probability of an IM on each of the FARs. We then apply the derived results in the proposed analytical framework to study the performance of the considered underlay cognitive molecular communication in which the secondary transmission is controlled to limit the co-channel interference at the primary receiver below a certain level.  The main contributions of this work are listed below.
	\begin{enumerate}
		\item We derive an approximate closed form expression for the hitting probability of an information molecule hitting each of the fully-absorbing receivers of different radii in a two-receiver system, which is an extension of the work done in \cite{Sabu2020a}.
		\item We derive an approximate equation for the hitting probability of a degradable information molecule hitting each of the fully-absorbing receivers of different radii in a two-receiver system.
		\item Based on the derived hitting probability expressions, we develop several important insights.
		\item We model a 3-D diffusion-limited cognitive molecular communication system with primary and secondary transmitter-receiver pairs. We consider an underlay strategy where the number of transmit molecules at the secondary transmitter is controlled to limit the co-channel interference at the primary receiver under a certain threshold.
		\item We derive the performance of the primary and the secondary links in terms of expected interference and bit error probability of both links. Several key insights are developed regarding optimal system parameters, gains due to transmit control strategy, and feasibility of cognition.
	\end{enumerate}
	The important symbols and notation used in this work is given in Table \ref{tab1}. 
\begin{table}[htbp]
	\caption{Notation Summary}
	\begin{center}
		\begin{tabular}{|p{.9in}|p{5in}|}
			\hline
			\textbf{Symbol}&\textbf{Definition} \\
			\hline
			$D$& Diffusion coefficient of the IMs in the propagation medium.\\
			\hline
			$\mu$& Degradation rate constant of the IM.\\
			\hline
			$\ts$& Bit duration.\\          
			\hline
			$\square_{\mathrm{P}},\ \square_{\mathrm{S}} $& Parameters related to the primary and secondary links respectively.\\
			\hline
			$\x_i,\ \y_i$& Locations of TX$_ i $ and FAR$_i $ in $ \mathbb{R}^3 $. $ i\in\{\pu,\su\} $\\
			\hline\vspace{0.01cm}
			$ \bar{i} $&\vspace{0.01cm} Denote compliment for $ i\in\{\pu,\su\} $. \ie $\bar{\su}=\pu $ and $ \bar{\pu}=\su $.\\
			\hline
			$r_\rsp$& Distance between TX$ _\su $ and the center of FAR$ _\pu $.\\
			\hline
			$ q_{bm} $&Probability of sending the bit $ \mathsf{b}\in\{0,1\} $ by the TX$_ m $.\\
			\hline
			$b_i[k]$& Information bit transmitted by the TX$_ i $ in the $ k $th time-slot.\\
			\hline
			$ \up[k] ,\newline \us[r_\rsp;k]$& Number of molecules emitted by primary and secondary transmitter in the $ k $th time-slot.\\          
			\hline	
			$ \ptwo{mi}{}{t,\mu} $ & Probability of an IM of degradation rate constant $ \mu $ emitted by the TX$_ m $ hitting on FAR$ _i $ within time $ t $ in the  presence of the other FAR.\\
			\hline	
			$ \ptwo{mi}{}{t} $ & Probability of an IM emitted by the TX$_ m $ hitting on FAR$ _i $ within time $ t $ in the  presence of the other FAR. Here $ \mu=0s^{-1} $.\\
			\hline	
			$\h{m}{i}[l-k]$ & Probability that a IM transmitted by TX$_ m $ in the slot $k\in\{1, 2,\cdots, l \}$ arrives at FAR$ _i $ in time-slot $l$.\\
			\hline	
			$ z_{mi} [l;k]$& Number of molecules received at the FAR$ _i $ in $ l $th slot due to the emission of molecules from the TX$_m $ at the $ k $th slot.\\
			\hline
			$ z_i[l]$& The total number of IMs observed at the FAR$ _i $ in the $ l $th time-slot.\\
			\hline
			$ \eta[i;l]$& Detection threshold at the FAR$ _i $ in the $ l $th time-slot.\\
			\hline
			$\pe{i,0}[l],\  \pe{i,1}[l]$& The probabilities of incorrect decoding for bit 0 and 1, respectively at the $ l $th time-slot.\\
			\hline
			$ \pe{i}[l]$& Total probability of error at the $ l $th time-slot.\\
			\hline
			
		\end{tabular}
		\label{tab1}
	\end{center}
\end{table}
	
	\section{System Model}\label{sec2}
		\begin{figure}
		\centering
		\includegraphics[width=\ifSC 0.7\else \fi\linewidth]{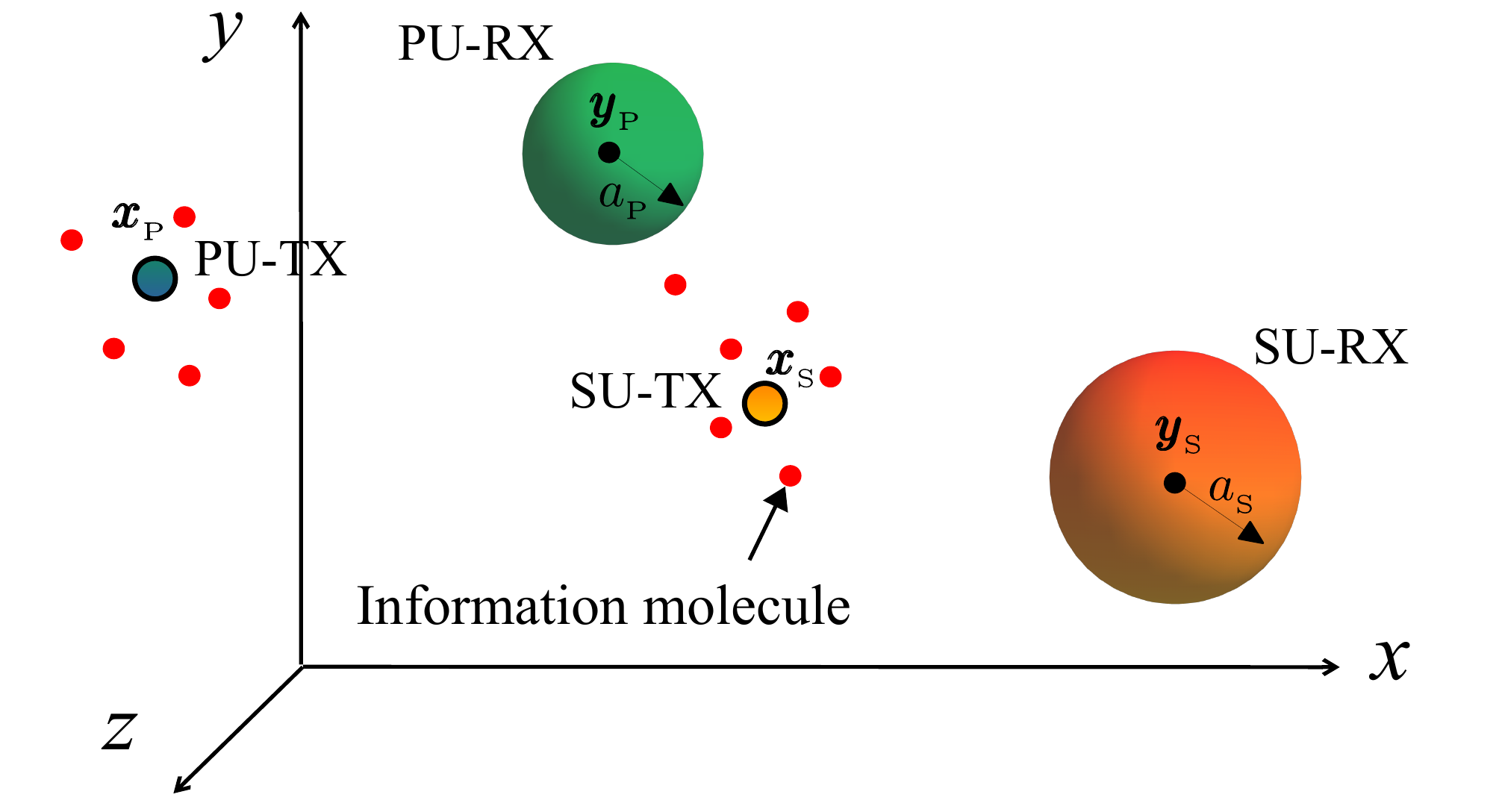}
		\caption{\small Schematic diagram of a 3-D diffusive cognitive molecular communication system where a secondary point transmitter is communicating with a secondary spherical FAR in presence of ongoing communication between primary point transmitter and spherical FAR.}
		\label{fig:BD2}
	\end{figure}  

This work considers a 3-D diffusion-limited scenario, with primary and secondary (cognitive) transmitter-receiver pairs, as shown in Fig.  \ref{fig:BD2}.  The primary and secondary links coexist in the communication medium with higher performance priority given to the primary link. The primary transmitter-receiver pair is unaware of the secondary transmission. In contrast, the secondary transmitter intelligently controls the number of transmit molecules so that the interference caused by the secondary transmitter on the primary receiver is below a threshold value in all time-slots. This strategy of communication is called \textit{underlay} cognitive-communication.
\subsection{Network Model}
The location of each of the communicating nodes is assumed to be fixed in the 3-D medium. Each transmitter is a point source, which can emit molecules into the propagation medium.  Each receiver is assumed to be spherical FAR, which absorbs all the molecules hitting its surface and counts them for decoding purpose. Let the primary point transmitter (TX$_\pu  $), primary FAR (FAR$_\pu  $), secondary point transmitter (TX$_\su  $)  and secondary FAR (FAR$_\su  $)  be located at positions $\x_{\p},\ \y_\p,\ \x_\s$ and $\y_\s$ respectively, in the $\rthree$ space. The FAR$_\pu  $ and FAR$_\su  $ are spherical fully-absorbing receivers of radius $ a_\p $ and $ a_\s $, respectively.  Both primary and secondary transmitter-receiver pairs use the same type of molecules as the carrier of information. Similar to several existing work such as \cite{Singhal2015a}, the transmitters and the receivers are assumed to be synchronized in time. 
\subsection{Modulation and Transmission Model}
Consider that the transmitters emit molecules at the beginning of a time-slot of duration $ \ts $. Let $\symp[l]$ and $\syms[l]$ denote the information bit transmitted by the TX$_\pu  $ and TX$_\su  $ respectively, at the $l$th time-slot. The bit $ b_m[l], m\in\{\mathrm{P},\mathrm{S}\} $ can be considered to be Bernoulli distributed with parameter $ q_{1m}$. We consider on-off keying (OOK) modulation for both transmissions. For primary communication, at the beginning of the time-slot, the TX$_\pu  $ emits a fixed number ($ \up[l]=N $) of molecules for a bit $ 1 $ and does not emit molecules for a bit $ 0 $. However, for secondary-communication, 
the TX$_\su  $ emits a variable  $ \us[r_\rsp;l]$  number of molecules for bit 1 and 0 molecules for bit 0 in any slot $ l $. For cognitive communication, the number of molecules $ \us[r_\rsp;l]$  is controlled according to system parameters (in particular, $ r_\rsp $ which is the distance between TX$ _\su $ and the center of FAR$ _\pu $. \ie $ \norm{\y_\p-\x_\s} $) to minimize the TX$_\su  $ interference at FAR$_\pu  $. It can also vary with $ l $ to have a better control on interference. The detailed derivation of $ \us[r_\rsp;l]$ considering the maximum and proportional interfering molecules constraints is given in Section \ref{s3}.
\subsection{Propagation Model}
We consider a pure diffusion-based propagation of IMs in this work. The IMs emitting from the TX$_\pu  $ and TX$_\su  $ travel in the communication medium via 3-D Brownian motion \cite{Woolard1928}. Further these IMs degrade over time with a degradation rate constant $ \mu $. Further, we consider that there are no potential collisions between the molecules when they propagate in the medium, and the molecules are immediately absorbed once they reach any one of the receivers. The diffusion coefficient for IMs is $ D $ which is assumed to be constant everywhere in medium.. 
\subsection{Channel Model} Let $ \ptwo{mi}{}{t,\mu} $ denote the probability of an IM (with degradation rate constant $ \mu $) emitted by the TX$ _m $  hitting on FAR$_i$  within time $ t $ in the  presence of the another FAR FAR$_\noti$ ($ m,i\in \{\mathrm{P},\mathrm{S}\} $). Let $\h{m}{i}[l-k]$ denote the probability that a IM transmitted by TX$_m$ in slot $k\in\{1, 2,\cdots, l \}$ arrives at FAR$ _i $ in time-slot $l$. Therefore,
\begin{align}
\h{m}{i}[l-k] =\ptwo{mi}{}{(l-k+1)\ts,\mu}-\ptwo{mi}{}{(l-k)\ts,\mu}. \label{FHTP}
\end{align}
\subsection{Receiver Observation}
Let us focus on the current time slot denoted by $ l $. Recall that, the transmit bit of the TX$ _m $ ($ m\in \{\pu,\su \} $) at the $ l $th time-slot ({\em{i.e.,} $b_m[l]$}) is Bernoulli distributed. 
\subsubsection{Observation at the {\normalfont FAR}$_\pu  $}
 Let us denote the number of received IMs at the FAR$_\pu  $ corresponding to the transmission of bit $\symp[l]\in\{0,1\}$ from the TX$_\pu  $ in the current time-slot as $z_\rpp[l;l]$. This quantity follows Binomial distribution with parameters $\symp[l]\up[l]$ and $h_\rpp[0]$ {\em i.e.,} $z_\rpp[l;l]\sim \mathcal{B}(\symp[l]\up[l],h_\rpp[0])$. This is due to fact that each IM's hit is a Bernoulli random variable with parameter $ h_\rpp[0] $, and each IM's propagation is independent of others.\par
 Due to Brownian motion, IMs emitted in previous $l-1$ time-slots also hits the receiver in the current and upcoming time-slots, resulting in inter-symbol interference (ISI). The number of stray molecules observed at the FAR$_\pu  $ in the $ l $th time-slot  due to the emission from the previous $k$th time-slot is denoted by $z_\rpp[l;k],\ 1\leq k\leq l-1$. Similar to $z_\rpp[l;l]$, $ z_\rpp[l;k] $ also follows Binomial distribution with parameters $ \symp[k]\up[k] $ and $ h_\rpp[l-k] $, {\em i.e.,} $z_\rpp[l;k] \sim\mathcal{B}\left(\symp[k]\up[k],h_\rpp[l-k]\right)$. Therefore, the total number of ISI molecules observed at the FAR$_\pu  $ at the $ l $th time-slot is 
$I_\rpp[l]=\sum_{k=1}^{l-1}z_\rpp[l;k]$. \par
Due to the presence of the secondary communication link, the IMs from the TX$_\su  $ also reaches the FAR$_\pu  $. Since both links are using same type of IMs, FAR$_\pu  $ cannot distinguish between them, resulting in co-channel interference (CCI) \cite{Kuran2012}. The number of interfering IMs reaching the FAR$_\pu  $  corresponding to the transmission of bit $\syms[k]$ by the TX$_\su  $ in the $k$th time-slot is denoted by $z_\rsp[l;k],\ 1\leq k\leq l$. Note that, $z_\rsp[l;k]$ is Binomial distributed with parameters $ \syms[k]\us[r_\rsp;k] $ and $ h_\rsp[l-k] $  {\em i.e.,} $z_\rsp[l;k] \sim\mathcal{B}\left(\syms[k]\us[r_\rsp;k],h_\rsp[l-k] \right)$.
The total CCI molecules observed at the FAR$_\pu  $ is 
$C_\rsp[l]=\sum_{k=1}^{l}z_\rsp[l;k]$.\par
Therefore, the total number of IMs observed at the FAR$_\pu  $ in the $ l $th time-slot is,
\begin{align}
 z_\p[l] = z_\rpp[l;l] + \underbrace{\sum_{k=1}^{l-1} z_\rpp[l;k]}_{\triangleq I_\rpp[l]}+ \underbrace{\sum_{k=1}^{l} z_\rsp[l;k]}_{\triangleq C_\rsp[l]}.\label{yp}
\end{align}	
\subsubsection{Observation at the {\normalfont FAR}$_\su  $}
Similarly, the total number of IMs observed at the FAR$_\su  $ is the sum of molecules observed at the receiver due to IMs emitted by the TX$_\su  $ in the current slot and past $ l-1 $ slots (ISI molecules), and IMs emitted by the TX$_\pu  $ in the current and previous slots (CCI molecules). \par
 The number of received IMs at the FAR$_\su  $ corresponding to the transmission of bit $\syms[l]\in\{0,1\}$ in the current $ l $th time-slot  is $z_\rss[l;l]$, where  $z_\rss[l;l]\sim \mathcal{B}(\syms[l]\us[r_\rsp;l],h_\rss[0])$. The total number of ISI molecules observed at the FAR$_\su  $ is $I_\rss[l]=\sum_{k=1}^{l-1}z_\rss[l;k]$, where $z_\rss[l;k] \sim\mathcal{B}\left(\syms[k]\us[r_\rsp;k],h_\rss[l-k]\right)$ and $\syms[k]$ is the bit transmitted by the TX$_\su  $ in previous $k$th time-slot. The total number of CCI molecules absorbed at the FAR$_\su  $ is $C_\rps[l]=\sum_{k=1}^{l}z_\rps[l;k]$, where $z_\rps[l;k] \sim\mathcal{B}\left(\symp[k]\up[k],h_\rps[l-k] \right)$.
\par
Therefore, the total number of IMs observed at the FAR$_\su  $ in the $ l $th time-slot is,
	\begin{align}
		z_\s[l] = z_\rss[l;l] + \underbrace{\sum_{k=1}^{l-1} z_\rss[l;k]}_{\triangleq I_\rss[l]}+ \underbrace{\sum_{k=1}^{l} z_\rps[l;k]}_{\triangleq C_\rps[l]}.\label{ys}
	\end{align}

\section{Transmit Control at the Secondary Transmitter}\label{s3}
We now describe the transmit control mechanism at TX$_\su  $ to maintain certain QoS for the primary link. Since the priority is given to the primary link, the interference from the secondary link must not affect the
performance of the primary link beyond a predefined limit. This is ensured by adapting the number of IMs emitted by TX$ _\su $ (\ie $ \us[r_\rsp;l] $) according to the channel between the FAR$ _\pu $ and TX$ _\su $ (which depends on $ r_\rsp $ only). Further, since the total interference also depends on the previous transmission due to ISI, $ \us[r_\rsp;l] $ may also need to be adjusted for each slot. Hence, $ \us[r_\rsp;l] $ will be a function of $ r_\rsp $ and the current slot index $ l $. In this section, we derive the expression for the number of molecules allowed to be emitted by the TX$ _\su $ in the $ l $th time-slot so that the interference caused at the FAR$ _\pu $ is below a desired threshold value  $u_\m$.\par
\begin{figure}
	\begin{subfigure}{0.5\textwidth}
		\centering
		\includegraphics[width=\linewidth]{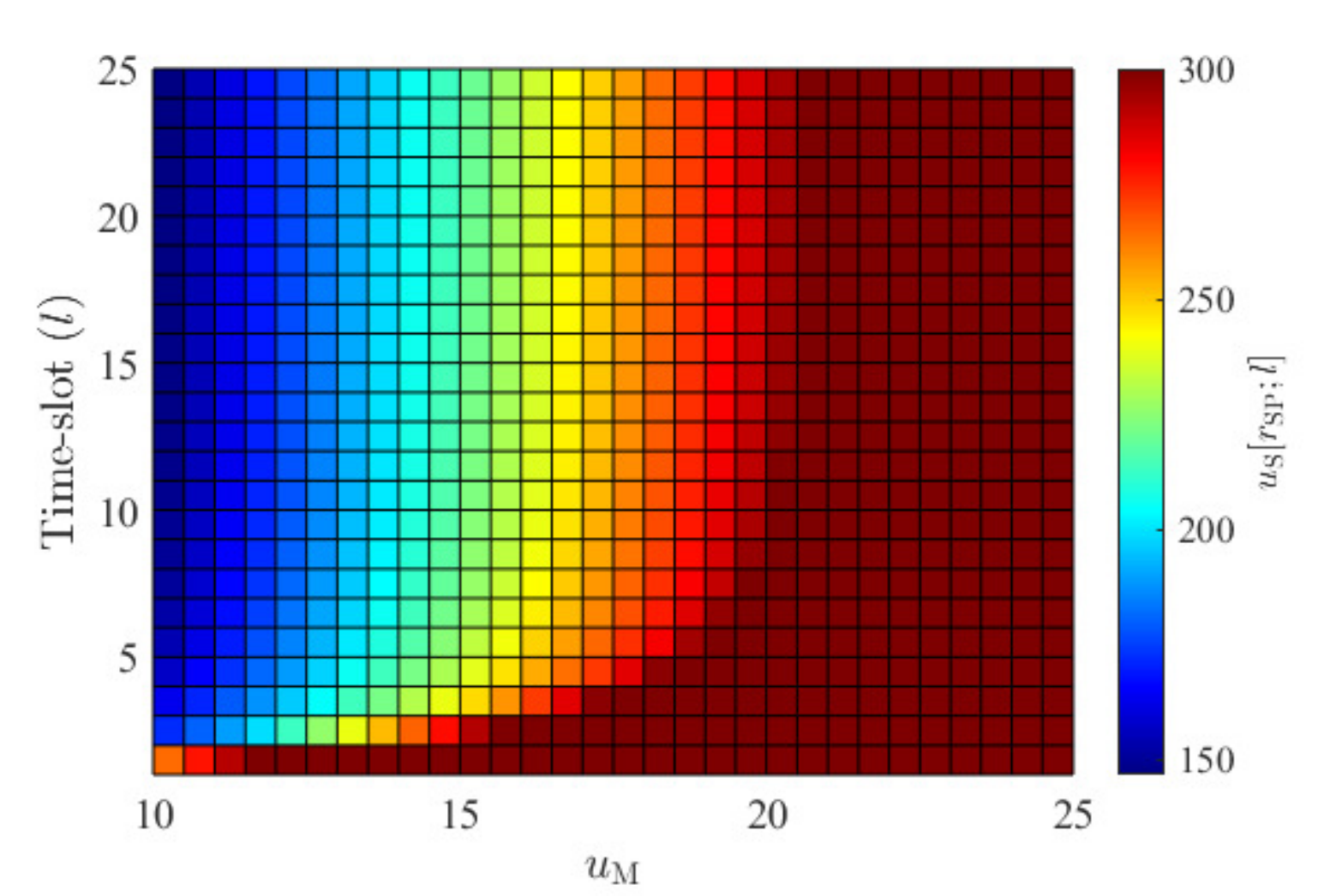}  
		\caption{$ \ts=1 $ s}
	\end{subfigure}
	\begin{subfigure}{0.5\textwidth}
		\centering
		\includegraphics[width=\linewidth]{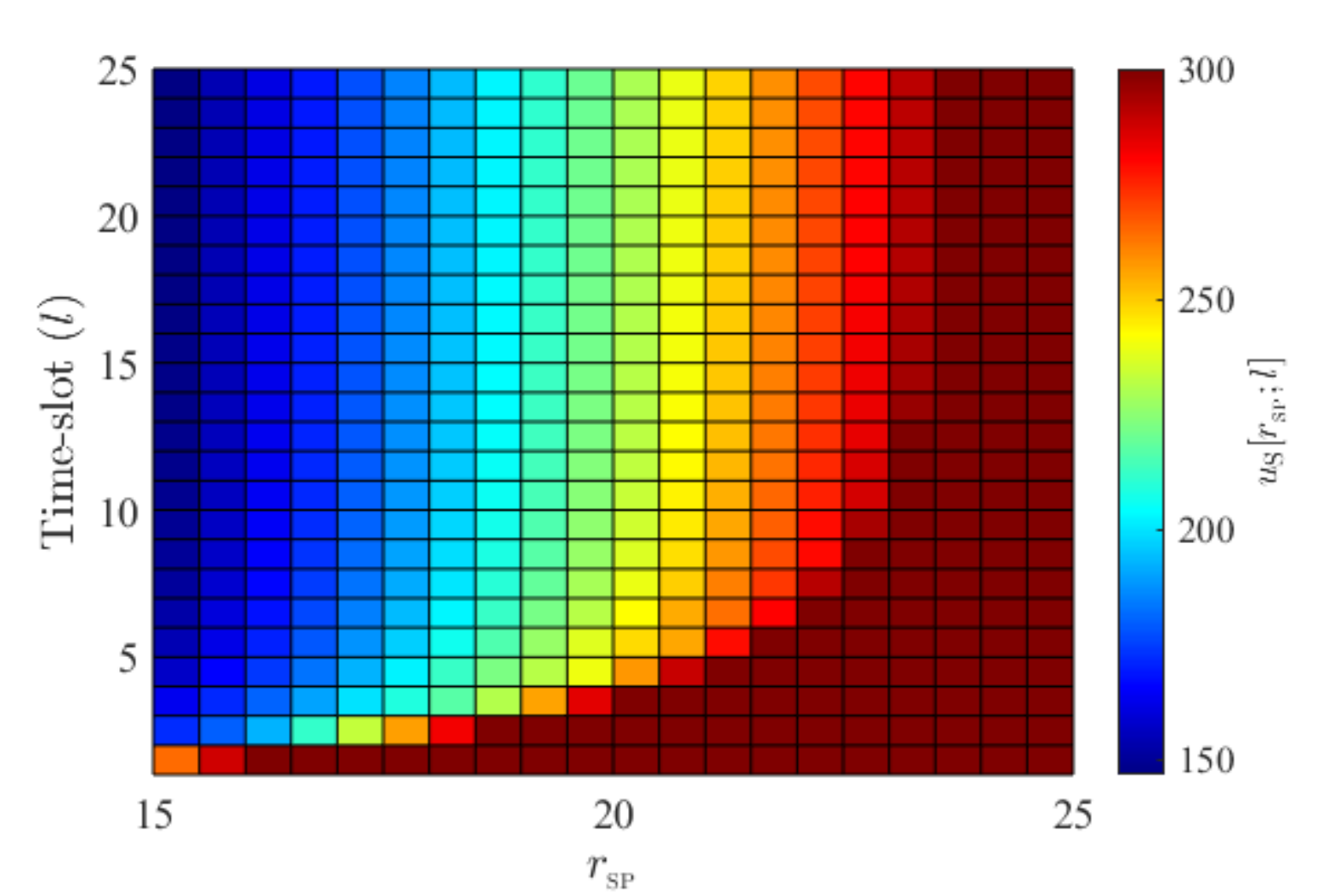}  
		\caption{$ \ts=1 $ s}
	\end{subfigure}
	\begin{subfigure}{\textwidth}
		\centering
		\includegraphics[width=0.5\linewidth]{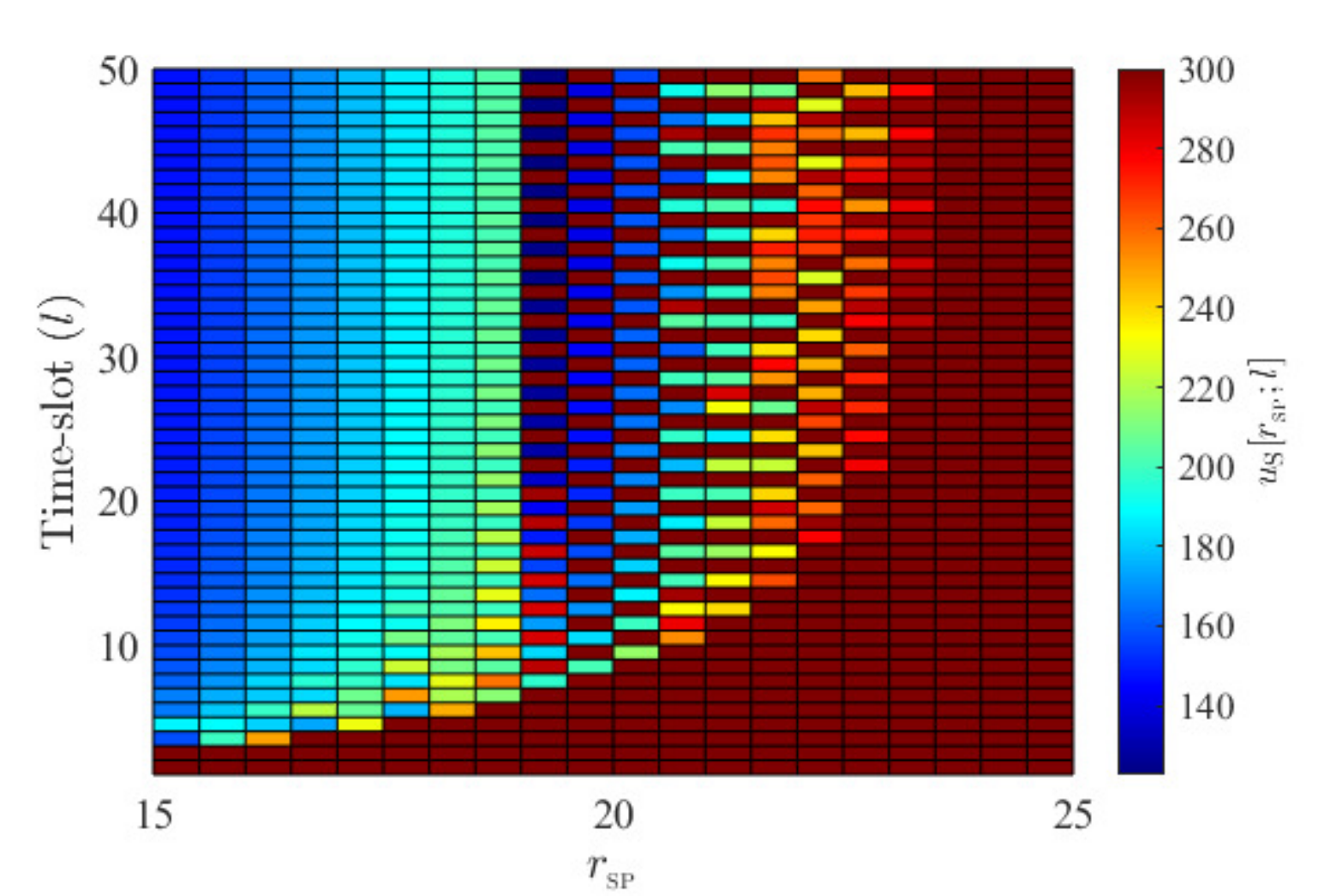}  
		\caption{$ \ts=0.4 $ s}
	\end{subfigure}
	
	\caption{(a) Variation of $\us[r_\rsp;l]$ with the threshold $u_\m$ and time-slot $l$ when$ \ r_\rsp=15\mu m$. (b) \& (c) Variation of $\us[r_\rsp;l]$ with  $r_\rsp$ and time-slot $l$ for different values of $ \ts $. Parameters: $D=100\mu m/s,\ u_\mathrm{L}=300$.}
	\label{fig:mkq}
\end{figure}

In other words, the transmitted number of molecules $ \us[r_\rsp;l] $ by the TX$ _\su $ is controlled such that the variation in $ r_\rsp $ and $ l $ does not result in the average number of interfering molecules
cross beyond an acceptable threshold $u_\m$ \ie
\begin{align}
	\expect{}{C_\rsp[l]} = \expect{}{\sum_{k=1}^{l} z_\rsp[l;k]} \leq u_\m. \label{N1}
\end{align}
The term $\expect{}{C_\rsp[l]}$ in \eqref{N1} can be further solved as 
\begin{align}
	\expect{}{C_\rsp[l]} = &\sum_{k=1}^{l} \expect{}{z_\rsp[l;k]}\nonumber\\
	=&\sum_{k=1}^{l} \pro\left[\syms[k]=1\right]\expect{}{z_\rsp[l;k]|\syms[k]=1}  +\pro\left[\syms[k]=0\right]\expect{}{z_\rsp[l;k]|\syms[k]=0}\nonumber \\
	=&\qpro{1}{\su} \sum_{k=1}^{l} \us[r_\rsp;k]h_\rsp[l-k]. \label{N2}
\end{align}
Further, due to the limited capacity of bio-nanomachines to generate a large number of molecules at a time, $ \us[r_\rsp;l] $ is upper bounded by the maximum number of molecules available to transmit, $ u_\mathrm{L}$. Now from \eqref{N1} and \eqref{N2}, the number of IMs allowed to be transmitted by TX$ _\su $ is given as
\begin{align}
	&\us[r_\rsp;l] =\floor*{\min\left\{u_\mathrm{L},\frac{1}{h_\rsp[0]} \left( \frac{u_\m}{\qpro{1}{\su}} - \sum_{k=1}^{l-1} \us[r_\rsp;k]h_\rsp[l-k]\right)^+\right\}},\label{ustran}
\end{align}
where $(x)^+ = \max(x,0)$. The transmitted number of molecules by the TX$_\su  $ is a function of $ r_\rsp $. Therefore, the TX$_\su  $ has to estimate the distance ($ r_\rsp $) \cite{Huang2013,Wang2015} in a practical implementation \cite{jingPowerControlISI2020}. Also, the TX$_\su  $ has to store the values of $ \us[r_\rsp;k], \ 1\leq k\leq l-1 $ at any time-slot $ l $. At large $ l $ (\ie steady state), $ \us[r_\rsp;l] $ may  become constant with time $ l $. 
\begin{remark}
	The lower bound on the number of transmitted molecules by the TX$_\su  $ at any time-slot $ l $ is zero. $ \us[r_\rsp;l] $ equals to zero when $\qpro{1}{\su}\sum_{k=1}^{l-1} \us[r_\rsp;k]h_\rsp[l-k]\geq u_\m $. That is, the number of transmitted molecules by the TX$_\su  $ at any time-slot $ l $ becomes zero when the average co-channel interference due to ISI is greater than the maximum allowed interference at the {\normalfont FAR}$_\pu  $ (\ie $ u_\m $). In this scenario, the secondary communication ceases while primary communication remains unaffected.
\end{remark}\label{rem2}
\begin{remark}
	For a system without ISI, the number of molecules transmitted by the cognitive transmitter in any time slot is
	\begin{align}
		\us[r_\rsp;l] =\floor*{\min\left\{u_\mathrm{L}, \frac{u_\m}{\qpro{1}{\su}h_\rsp[0]}\right\}}.\label{erm2}
	\end{align}
Systems with large time-slot duration and/or fast molecule degradation have negligible number of ISI molecules and hence \eqref{erm2} will be valid for them also.
\end{remark}
\begin{remark}\label{r4}
	At the steady state (large $ l $), the number of molecules transmitted by the secondary transmitter in each time slot becomes constant and is  given as
	\begin{align}
		&\us[r_\rsp;\infty]\leq\floor*{\min\left\{u_\mathrm{L},\frac{u_\m}{\qpro{1}{\su}\ptwo{\s\p}{}{\infty,\mu}} \right\}}.\label{eqsteady}
	\end{align}
\end{remark}
\begin{IEEEproof}
	See Appendix \ref{steady}.
\end{IEEEproof}
 Note that, $ \us[r_\rsp;l] $ can be oscillatory with $ l $ also. For example, if $ \ts $ is too small, the received number of molecules at the current time slot can be much less than the previous slot, causing $ \us[r_\rsp;l] $ to rise and fall in the alternative time-slots. The steady state and oscillatory behavior of $ \us[r_\rsp;l] $ is further discussed with the help of Fig. \ref{fig:mkq}. \par
Fig. \ref{fig:mkq} (a) shows the variation of $\us[r_\rsp;l]$ with the interference threshold $u_\m$ at FAR$_\pu  $, and time-slot $l$. When $u_\m$ decreases, the allowed interference at the FAR$_\pu  $ due to TX$_\su  $ reduces, and $\us[r_\rsp;l]$ is decreased. Also, with the increase in the $ l $, the ISI caused by the TX$_\su  $ to the FAR$_\pu  $ increases, and the transmitted number of molecules  $\us[r_\rsp;l]$ by the TX$_\pu  $ is reduced to limit the interference by $ u_\m $. When $ r_\rsp $ is small, the TX$_\su  $ is near to FAR$_\pu  $, resulting in high co-channel interference, and $\us[r_\rsp;l]$ is reduced to limit it.  When $ r_\rsp $ is large, the TX$_\su  $ is far from FAR$_\pu  $, resulting in low co-channel interference and $\us[r_\rsp;l]$ can be high. This trend can be seen in Fig. \ref{fig:mkq} (b). Also in Fig. \ref{fig:mkq} (a) and (b), $ \us[r_\rsp;l] $ reaches a steady state value when $ l $ is increased. $ \us[r_\rsp;l] $ can be oscillatory also if $ \ts $ is too small, as seen in Fig. \ref{fig:mkq} (c).\par
Now to proceed further, we require $ \ptwo{mi}{}{t,\mu}  $ \ie the fraction of IM emitted by TX$ _m $ that are hitting the FAR$ _i $ in the presence of FAR$ _\noti $ until time $ t $ while considering the degradation of IMs which we do next.
	\section{Hitting Probability of an IM}\label{sec3}
			\begin{figure}[!h]
		\centering
		\includegraphics[width=\ifSC 0.7\else \fi\linewidth]{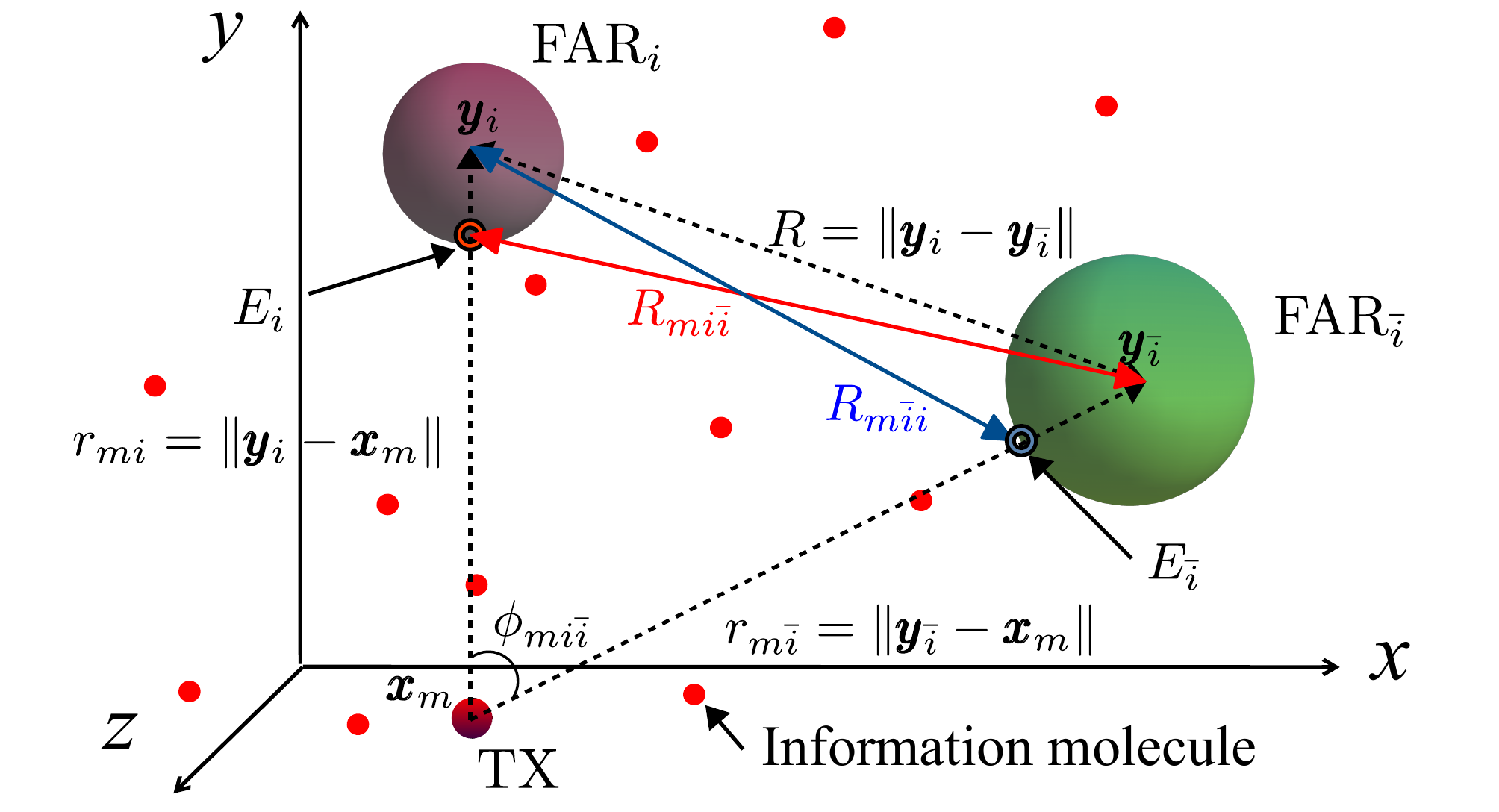}
		\caption{\small Schematic diagram of a 3-D diffusion based cognitive molecular communication system with a point transmitter and two spherical FARs.}
		\label{fig:hp}
	\end{figure}  
We now analyze a system with one point transmitter (located at $ \x_m $) and two FARs (located at $ \y_i $ and $ \y_\noti,\ i\in\{\pu,\su\} $) of unequal size (see Fig. \ref{fig:hp}) to derive an approximate expression for hitting probability of an IM at one of the FAR. We first consider the case with no molecular degradation and then extend the analysis for the case when molecular degradation is present. Note that an exact equation for a 3-D MCvD system with multiple fully-absorbing receivers is not available in the literature. The derived equation in this work is different than the one derived in \cite{Sabu2020a}, which considered FARs of the same size only.
	\subsubsection{Without molecular degradation}
	Let $ \ptwo{mi}{}{t}=\ptwo{mi}{}{t,\mu=0} $, $m, i\in\{\mathrm{P},\mathrm{S}\}$, denote the fraction of IMs (without degradation) absorbed within time $t$ by the FAR located at $\y_{i}$ in the presence of another FAR at $ \y_\noti $ when the transmitter is located at $\x_{m}$. This quantity is equal to the probability of an IM being absorbed by the FAR at $ \y_i $ in the presence of other FAR at $ \y_j $ until time $t$.
	 \begin{theorem}\label{t1}
		The closed-form approximate expression for $  \ptwo{mi}{}{t} $ can be derived as
	\begin{align}
	\ptwo{mi}{}{t}=&\sum_{n=0}^{\infty}\left(\frac{a_ia_{\noti}}{R_{mi\noti}R_{m\noti i}}\right)^{n}\left[\frac{a_i}{r_{mi}}
	\erfc\left(\frac{\Phi_{mi}(n)}{\sqrt{4Dt}}\right)\right.
	-\left.\frac{a_ia_{\noti}}{r_{m\noti}R_{m\noti i}}
	\erfc\left(\frac{\Psi_{mi}(n)}{\sqrt{4Dt}}\right)\right],\label{ndeg}
	\end{align}
	where $\Phi_{mi}(n)= r_{mi}-a_i+n(R_{m\noti i}{-}a_i)+n(R_{mi\noti}{-}a_{\noti}),\ \Psi_{mi}(n)= r_{m\noti}-a_{\noti}+(n+1)(R_{m\noti i}{-}a_i)+n(R_{mi\noti}{-}a_{\noti}) $ with $r_{mj}= \norm{\y_{j}-\x_{m}},\ j\in\{i,\noti\} $ denoting the distance between the transmitter located at $ \x_m $ and the intended FAR located at $ \y_j $, and $R_{mi\noti}^2=(r_{mi}-a_i)^2+r_{m\noti}^2-2(r_{mi}-a_i)r_{m\noti}\cos(\phi_{mi\noti})$ denoting the square distance between the center of FAR$ _\noti $ and the point at the surface of FAR$ _i $, which closest to the transmitter ($ E_i $). Here, $ \phi_{mi\noti}$ is the angle between the centers of FAR$ _i $ and FAR$ _\noti $ when viewed from $ \x_m $.
	\end{theorem}
\begin{IEEEproof}
	See Appendix \ref{app:A}.
\end{IEEEproof}

 It is important to note that the error arising due to this approximation is small if (i) the distance between the TX$_m$ and each FAR is significantly larger than the
	radius of the FAR i.e., $\norm{\y_{i}-\x_{m}}\gg a_i$  and $\norm{\y_\noti-\x_{m}}\gg a_{\noti}$, and (ii)
	the distance between the FARs is significantly
	larger than the maximum of the radius of FARs i.e., $R=\norm{\y_{i}-\y_\noti} \gg \max{\left(a_\p,a_\s\right)} $ as verified in Fig. \ref{fig:error}.\par 
	\begin{coro}
		The fraction of IMs with eventually hitting FAR$_i$ is
		\begin{align}
		\ptwo{mi}{}{\infty}=\frac{a_iR_{mi\noti}}{R_{mi\noti}R_{m\noti i}-a_ia_{\noti}}\left[\frac{R_{m\noti i}}{r_{mi}}-\frac{a_{\noti}}{r_{m\noti}}\right]\label{cor1_1}.
		\end{align}
	\end{coro}
	The hitting rate of molecules on the FAR located at $\y_{i}$ in the presence of other FAR at time $ \tau $ can be derived by taking the derivative of \eqref{ndeg} with respect to $ t $. That is,
	\begin{align}
	\hr{mi}{\tau}=\sum_{n=0}^{\infty}\left(\frac{a_ia_{\noti}}{R_{mi\noti}R_{m\noti i}}\right)^{n}&\left[\frac{a_i}{r_{mi}}
		\frac{\Phi_{mi}(n)}{\sqrt{4\pi D \tau^3}}\exp\left( -\frac{\Phi_{mi}(n)^2}{4D\tau}\right)\right.\nonumber\\
		&-\left.\frac{a_ia_{\noti}}{r_{m\noti}R_{m\noti i}}
		\frac{\Psi_{mi}(n)}{\sqrt{4\pi D \tau^3}}\exp\left( -\frac{\Psi_{mi}(n)^2}{4D\tau}\right)\right].\label{kappaij}
	\end{align}
	\subsubsection{With molecular degradation} 
	We now consider degradable IMs with exponential degradation where the probability that a molecule will not degrade over time  $t$ is, $\exp{\left(-\mu t\right)}$. Here, $\mu$ denotes the reaction rate constant of molecular degradation which is related to the half-time as $\mu=\ln(2)/\Lambda_{1/2}$. When the reaction rate tends to zero ($\mu\rightarrow 0$, {\em i.e. } half-time is infinity $\Lambda_{1/2}\rightarrow \infty$), the molecule will never undergo degradation.
	We also assume that the molecule does not get involved in any other reactions. \par
	Now, using the exponential distribution for molecular life expectancy, the fraction of non-degraded IMs reaching the FAR$_i$ in the presence of the other FAR within time $ t $, due to the emission of IMs from the transmitter at $\x_{m}$ is given by
	\begin{align}
	\ptwo{mi}{}{t,\mu}
	=&\int_{0}^{t}\hr{mi}{\tau} \exp\left(-\mu \tau\right)\dd \tau.\label{edeg}
	\end{align}
	Further, substituting \eqref{kappaij} in \eqref{edeg}, the closed-form approximate expression for hitting probability can be derived as presented in Theorem \ref{t2} below.
	\begin{theorem}\label{t2}
		The probability of a non-degraded IM hitting the FAR located at $\y_{i}$ within time $ t $, due to the emission from the transmitter at $\x_{m}$ in the presence of the other FAR is
		\begin{align}
		&\ptwo{mi}{}{t,\mu}=
		\sum_{n=0}^{\infty}\left(\frac{a_ia_{\noti}}{R_{mi\noti}R_{m\noti i}}\right)^{n}\left[\frac{a_i}{r_{mi}}f(\Phi_{mi}(n))
		-\frac{a_ia_{\noti}}{r_{m\noti}R_{m\noti i}}
		f(\Psi_{mi}(n))\right],\label{deg}
		\end{align}
		where 
		\begin{align*}
			f(x)=&\frac12\left[\erfc\left(\frac{x}{\sqrt{4Dt}}{+}\sqrt{\mu t}\right)\exp\left(x\sqrt{\frac{\mu}{D}}\right)\right.
			\left.+\erfc\left(\frac{x}{\sqrt{4Dt}}{-}\sqrt{\mu t}\right)\exp\left({-}x\sqrt{\frac{\mu}{D}}\right)\right] .
		\end{align*} 
	\end{theorem}
\begin{IEEEproof}
	See Appendix \ref{app:deg}.
\end{IEEEproof}
	\begin{coro}
		$\lim\limits_{\mu\rightarrow 0}\ptwo{mi}{}{t,\mu}=\ptwo{mi}{}{t}$.
	\end{coro}

	\begin{coro}
		The fraction of IMs eventually hitting FAR$_i$ is
		\begin{align}
		&\ptwo{mi}{}{\infty,\mu}=
		\frac{a_iR_{mi\noti}}{R_{mi\noti}R_{m\noti i}-a_ia_{\noti}\exp\left(-\left(R_{mi\noti}+R_{m\noti i}-a_i-a_{\noti}\right)\sqrt{\frac{\mu}{D}}\right)}\nonumber\\
		&\spc\spc\spc\times\left[\frac{R_{m\noti i}}{r_{mi}}\exp\left(-\left(r_{mi}-a_i\right)\sqrt{\frac{\mu}{D}}\right)\right.
		\left.
		-\frac{a_{\noti}}{r_{m\noti}}
		\exp\left(-\left(r_{m\noti}-a_{\noti}+R_{m\noti i}-a_i\right)\sqrt{\frac{\mu}{D}}\right)\right]\label{cor2}.
		\end{align}
		\begin{IEEEproof}
			When $ t\rightarrow\infty $, \eqref{deg} gives
			\begin{align}
	\ptwo{mi}{}{\infty,\mu}=&\sum_{n=0}^{\infty}\left(\frac{a_ia_{\noti}}{R_{mi\noti}R_{m\noti i}}\right)^{n}\left[\frac{a_i}{r_{mi}}\exp\left(-\Phi_{mi}(n)\sqrt{\frac{\mu}{D}}\right)\right.
	\left.
	-\frac{a_ia_{\noti}}{r_{m\noti}R_{m\noti i}}
	\exp\left(-\Psi_{mi}(n)\sqrt{\frac{\mu}{D}}\right)\right].\label{cor2_1}
			\end{align}
			Simplifying \eqref{cor2_1} further gives \eqref{cor2}.
		\end{IEEEproof}		
	\end{coro}	
\subsection{Validation of Hitting Probability Equation of a System With Two FARs}

\begin{figure}
	\centering
	\includegraphics[width=0.6\linewidth]{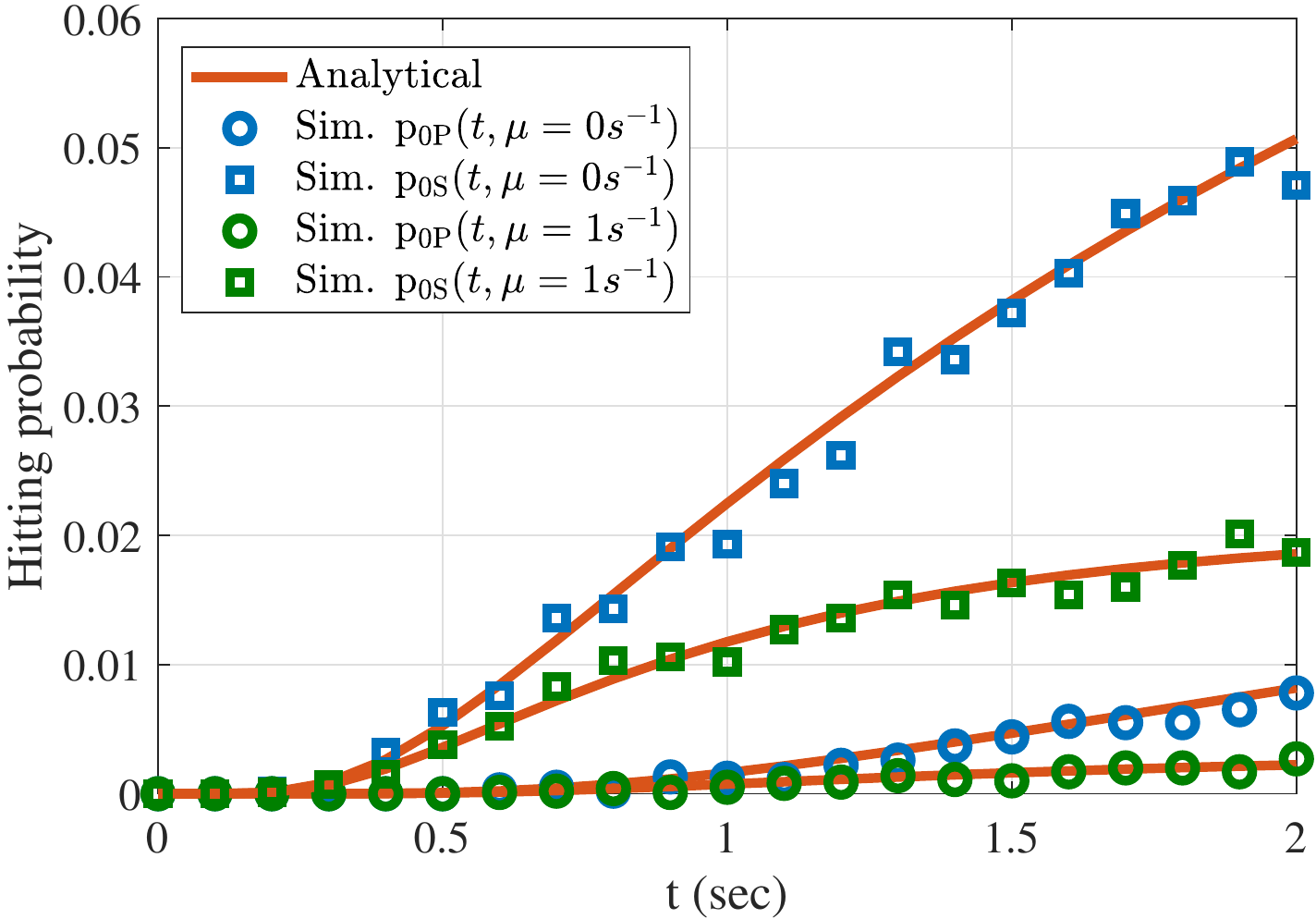}
	\caption{\small{Hitting probability versus time for different degradation rate constant. Parameters: $D=100\mu m/s,\ \x_m=[0,0,0],\ \y_\p=[-30,-20,0],\ \y_\s=[25,10,0],\ D=100\mu m/s,\ a_\p=3\mu m,\ a_\s=5\mu m,\ \Delta t=10^{-4}s. $}}
	\label{fig:ch}
\end{figure}
\begin{figure}
	\centering
	\includegraphics[width=0.6\linewidth]{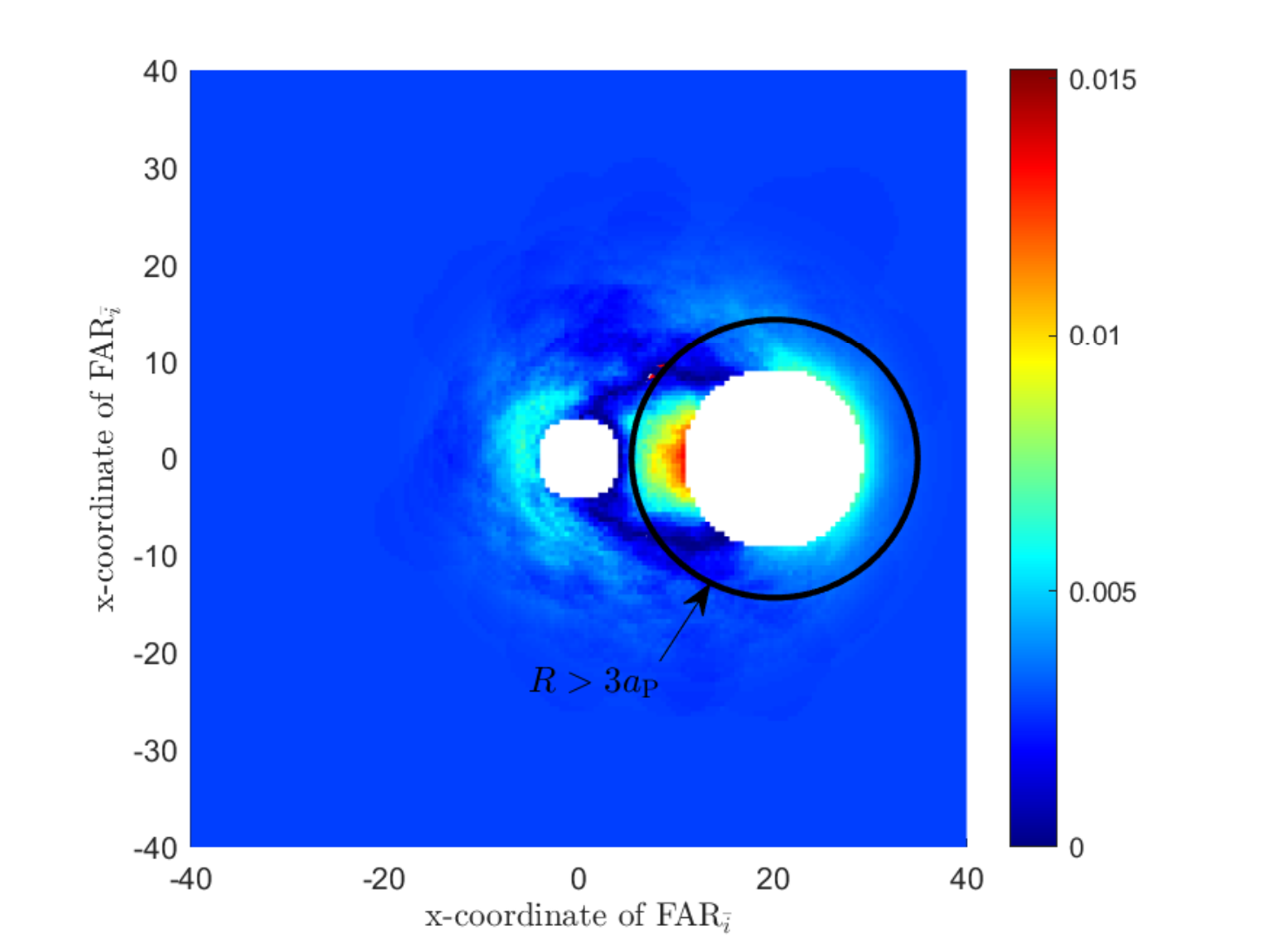}
	\caption{\small{Approximation error in the hitting probability of IM on FAR$ _i $ located at $ \y_i=[20,0,0] $ in the presence of FAR$_\noti $ in $ \rthree $. Parameters: $ \x_m=[0,0,0],\ D=100\mu m/s,\ a_\p=5\mu m,\ a_\s=4\mu m,\ \Delta t=10^{-4}s,\ t=1s$.}}
	\label{fig:error}
\end{figure}
The approximate analytical expressions \eqref{ndeg} and \eqref{deg} derived are validated using particle based simulations in Fig. \ref{fig:ch}. The step size $ \Delta t $ chosen for the simulation is $ 10^{-4} $s. The transmitter is assumed to be located at the origin, \ie  $\ \x_m=[0,0,0] $. The FAR$_\pu  $ is located at $ \y_\p=[-30,-20,0] $ and the FAR$_\su  $ is located at $  \y_\s=[25,10,0] $. It is evident from the Fig. \ref{fig:ch} that, the derived approximate analytical expressions considering and without considering molecular degradation are in good match with the particle-based simulation results for the chosen parameters.\par
The accuracy of approximating the hitting point of an IM on FAR$ _i $ by $ E_i $ can be observed in Fig. \ref{fig:error}, which shows the absolute error ($ |\text{Exact value-Analytical value}| $) when FAR$_i  $ is at a fixed location, while FAR$ _\noti $ is moved in $ \rthree $. It can be seen that the absolute error is negligible when $ R=\norm{\y_i-\y_\noti} >3a_\p$. The low values of absolute error guarantee the accuracy of the derived equations. \par
The derived approximate hitting probability expressions are used to model and analyze the underlay cognitive molecular communication system discussed in the upcoming sections.
We now incorporate this derived result to evaluate the performance of the cognitive system.

		\section{Expected number of absorbed molecules}
		In this section, we derive the expected number of molecules absorbed at the FAR$_\pu  $ and FAR$_\su  $ at the $ l $th time-slot.\par

	The total number of molecules absorbed at the FAR$_\pu  $ on $ l $th slot is $ z_\p[l] $. Therefore the expected total number of molecules absorbed at the FAR$_\pu  $  is
	\begin{align}
	\expect{}{z_\p[l]}&=N\qpro{1}{\pu}\sum_{k=1}^{l}h_\rpp[l-k]+\qpro{1}{\su}\sum_{k=1}^{l}\us[r_\rsp;k]h_\rsp[l-k]\nonumber\\
	&=N\qpro{1}{\pu}\ptwo{\rpp}{}{l\ts,\mu}+\qpro{1}{\su}\sum_{k=1}^{l}\us[r_\rsp;k]h_\rsp[l-k].
	\end{align}
	Similarly, the total number of molecules absorbed at the FAR$_\su  $ on $ l $th slot is $ z_\s[l] $. Therefore the expected total number of molecules absorbed at the FAR$_\su  $  is
	\begin{align}
	\expect{}{z_\s[l]}=N\qpro{1}{\pu}\ptwo{\rps}{}{l\ts,\mu}+\qpro{1}{\su}\sum_{k=1}^{l}\us[r_\rsp;k]h_\rss[l-k].
	\end{align}
\begin{figure}
	\centering
	\includegraphics[width=0.6\linewidth]{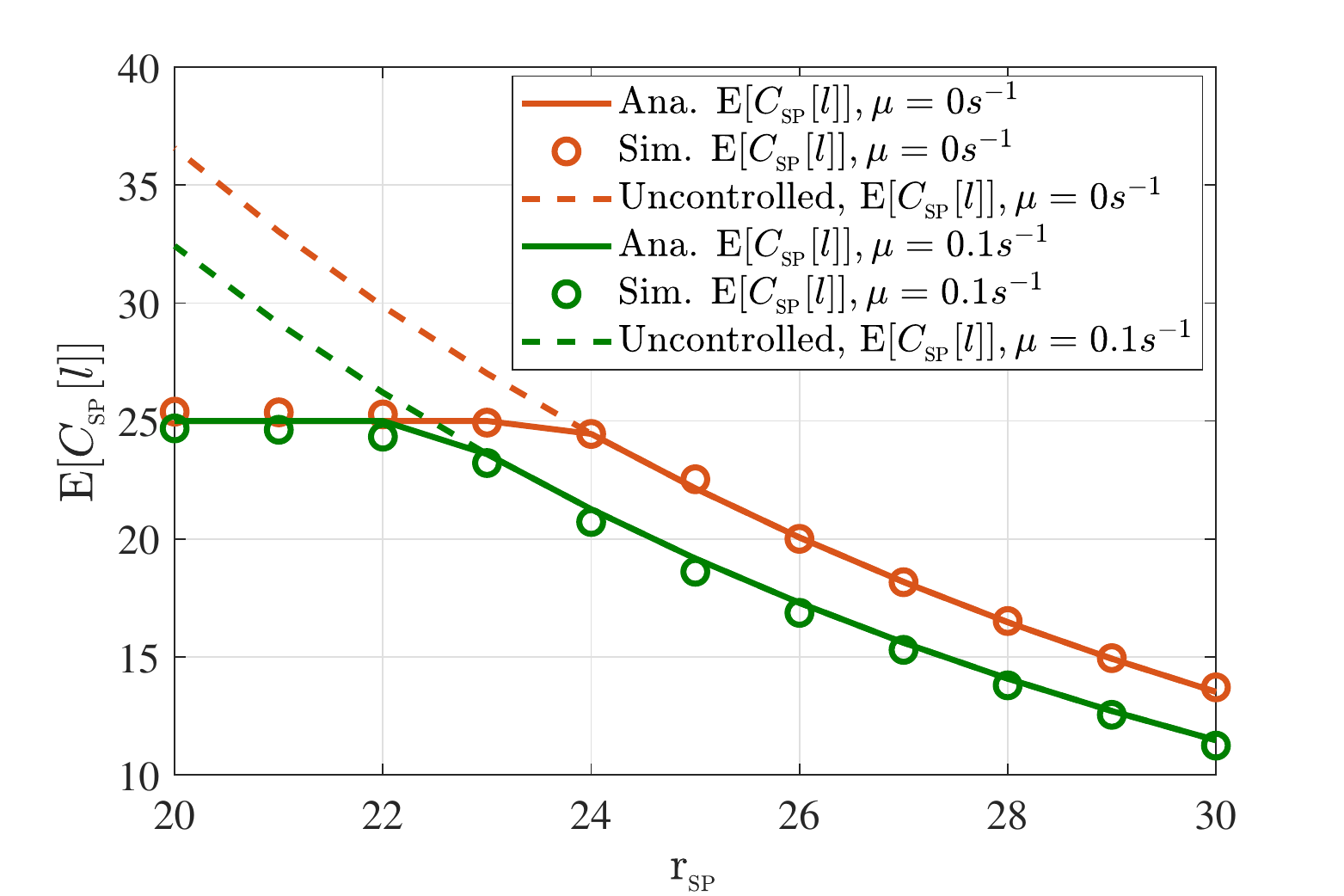}
	\caption{\small Variation of the expected number of absorbed molecules at the FAR$_\pu  $ due to the emission from TX$_\su  $ with respect to $r_\rsp$. Initial position of TX$_\su  $ is $ \x_\s=[0, 0, 0] $. TX$_\su  $ is shifted in positive x axis direction. Parameters: $D=100\mu m/s,\ l=3, u_\m=25, u_L=1000,\x_\p=[55, 0, 0],\ \y_\p=[30, 0, 0],\ \y_\s=[30, 50, 0] $.}
	\label{fig:expectedrand}
\end{figure}
Fig. \ref{fig:expectedrand} shows the variation of the expected number of absorbed molecules at the FAR$_\pu  $ due to the emission from TX$_\su  $ ($ \expect{}{C_\rsp[l]} $) with respect to $r_\rsp$. When $r_\rsp$ is small, the interference from the TX$_\su  $ to the FAR$_\pu  $ is high, and $\us[r_\rsp;l]$ is adapted (as in eq. \eqref{ustran}) at the TX$_\su  $ to limit the allowed interference to $ u_\m $ molecules at FAR$_\pu  $. When $r_\rsp$ increases, the expected number of interfering molecules observed at the FAR$_\pu  $ falls below $ u_\m $ due to the lossy channel. If $\us[r_\rsp;l]$ is uncontrolled ($ \us[r_\rsp;l]=1000 $, represented by dashed lines in Fig. \ref{fig:expectedrand}.) the co-channel interference at the FAR$_\pu  $ keep on increasing when $r_\rsp$ decreases. Also, when the molecular degradation rate increases, $ \expect{}{C_\rsp[l]} $ falls due to the degradation of molecules.\par
In the following section, we derive the novel closed-form expressions for the probability of bit error to analyze the performance at FAR$_\pu  $ and FAR$_\su  $.
	\section{Analysis of the Probability of Bit Error} \label{sec4}
	At the end of $ l $th time-slot, the number of IMs absorbed ($z_i[l],\ i\in \{\pu,\su\}$) at each FAR is compared with the threshold $\eta[i;l] $ for decoding the bit $\symi[l]$ transmitted by the TX$ _i $  at the beginning of $l$th time-slot. If $ z_i[l]<\eta[i;l]$, then the transmitted bit $\symi[l]$ is estimated as $\esymi[l]=0$, otherwise $\esymi[l]=1$. An error occurs when the transmitted bit $\symi[l]=0$ is decoded as $\esymi[l]=1$ and vice versa. Thus, the total probability of bit error ($\pe{i}$) is given by
	\begin{align}
	\pe{i}[l]= &\pro\left[\symi[l]=0\right] \pe{i,0}[l]+\pro\left[\symi[l]=1\right] \pe{i,1}[l],\label{pei}
	\end{align}
	where $\pe{i,0}[l]$  and $ \pe{i,1}[l], i\in\{\pu,\su\}$ are the probabilities of incorrect decoding for bit 0 and 1, respectively, and are defined as
	\begin{align}
	\pe{i,0}[l]&=\pro\left[\esymi[l]=1\mid\symi[l]=0\right],\nonumber\\
	\pe{i,1}[l]&=\pro\left[\esymi[l]=0\mid\symi[l]=1\right].
	\end{align}
		To derive the closed-form expression for $\pe{i,0}[l]$ and $\pe{i,1}[l]$, we employ the moment generating function approach and the results are presented in Theorem \ref{th2}.  
	\subsection{Probability of Bit Error at {\normalfont FAR}$_\pu  $ and {\normalfont FAR}$_\su  $}
	\begin{theorem}\label{th2}
		The probability of bit
		error at the receiver  $ i\in\{\pu,\su\} $ given by \eqref{pei} with the probability of incorrect
		decoding for bit 1 and 0 given as
		\begin{align}
		\pe{i,1}[l]=&\sum_{n=0}^{\eta[i;l]-1}\sum_{\substack{n_0+n_1+\cdots+n_{2l-1}=n\\ n_s<u_\noti(2l-s),\ \forall l\leq s\leq 2l-1}}\!\!\frac{1}{n_0!n_1!\cdots n_{2l-1}!}
		\beta_i^{(n_0)}(l)\prod_{r=1}^{l-1}\alpha_{ii}^{(n_r)}(r)\prod_{s=l}^{2l-1}\alpha_{i\noti}^{(n_s)}(s-l)\label{Ap},
		\end{align}
		and
		\begin{align}
		\pe{i,0}[l]=1-&\sum_{n=0}^{\eta[i;l]-1}\sum_{\substack{n_1+n_2+\cdots+n_{2l-1}=n\\ n_s<u_\noti(2l-s),\ \forall l\leq s\leq 2l-1}}\frac{1}{n_1!n_2!\cdots n_{2l-1}!}
		\prod_{r=1}^{l-1}\alpha_{ii}^{(n_r)}(r)\prod_{s=l}^{2l-1}\alpha_{i\noti}^{(n_s)}(s-l)\label{Bp},
		\end{align}
		where the second sum extends over all $ n_0,n_1,\cdots ,n_{2l-1} $ of non-negative integers such that $ n_0+n_1+\cdots+n_{2l-1}=n $. The variables $\alpha_{ij}^{(k)}(s)$ for $\ j\in\{i,\noti\}$ and $\beta_i^{(k)}(l)$ are defined as,
				\begin{align}
\alpha_{ij}^{(k)}(s)=&\qpro{0}{j}\mathbbm{1}(k=0)+\qpro{1}{j}\left(1-h_{ji}[s]\right)^{u_j(l-s)-k} 
\times h_{ji}[s]^{k}\frac{u_j(l-s)!}{\left(u_j(l-s)-k\right)!},\label{aij1}\\ 
\text{and}\nonumber\\
\beta_i^{(k)}(l)=&\frac{1}{\qpro{1}{i}}\left(\alpha_{ii}^{(k)}(l)-\qpro{0}{i}\mathbbm{1}(k=0)\right),
		\end{align}
where $ u_\p(s)=N $ and $ u_\s(s)=u_\s\left[r_\rsp;s\right] $ (with some abuse of notation). 
	\end{theorem}
\begin{IEEEproof}
	See Appendix \ref{app:B}.
\end{IEEEproof}
\begin{remark}
Consider the {\normalfont FAR}$_\pu  $ probability of bit error performance with the variation in the distance between {\normalfont FAR}$_\pu  $ and TX$_\su  $. When the {\normalfont TX}$_\su  $ moves closer to the {\normalfont FAR}$_\pu  $, 	$ h_\rsp[s-l]$ in \eqref{aij1} increases and $ \pe{\p}[l] $ increases due to the increase in $ \alpha_{i\noti}^{(k)}(s) $ if the transmitted molecules at the TX$_\su  $ is not controlled. However, for in controlled transmission, an increase in $ h_\rsp[s-l] $ is counteracted by a corresponding decrease in $ \us[r_\rsp;2l-s]$ to reduce the rise in $ \pe{\p}[l] $.
\end{remark}
	\begin{coro}
		Assuming $ \eta=1 $ and the probability of sending bit $ 0 $ and $ 1 $ as $ 1/2 $, the total probability of error at the $ i $th receiver is
	\begin{align}
	\pe{i}[l]=0.5\left(1+\prod_{r=1}^{l-1}a_{ii}^{(0)}(r)\prod_{s=l+1}^{2l}\alpha_{i\noti}^{(0)}(s-l)\left[\beta_i^{(0)}(l)-1\right] \right).\label{cpei}
\end{align}
From the above equation, we can verify that the second term with product terms is negative and with the increase in ISI, the value of the second term reduces and $ \pe{i}[l] $ increases.
\end{coro}
\begin{coro}
For a system without ISI, the probability of bit
error at the FAR$_i,\ i\in\{\pu,\su\} $ is given by 
\begin{align}
	\pe{i}[l]=&\qpro{1}{i}\sum_{n=0}^{\eta[i;l]-1}\frac{1}{n!}\sum_{k=0}^n\!\!{n\choose k}\beta_i^{(n-k)}(l)\alpha_{i\noti}^{(k)}(l)
	+\qpro{0}{i}\left[1-\sum_{n=0}^{\eta[i;l]-1}\frac{1}{n!}\alpha_{i\noti}^{(n)}(l)\right]\label{noisi}.
\end{align}
\end{coro}
\subsection{Effect of Threshold on Probability of Bit Error}
\begin{figure}
	\centering
	\includegraphics[width=0.6\linewidth]{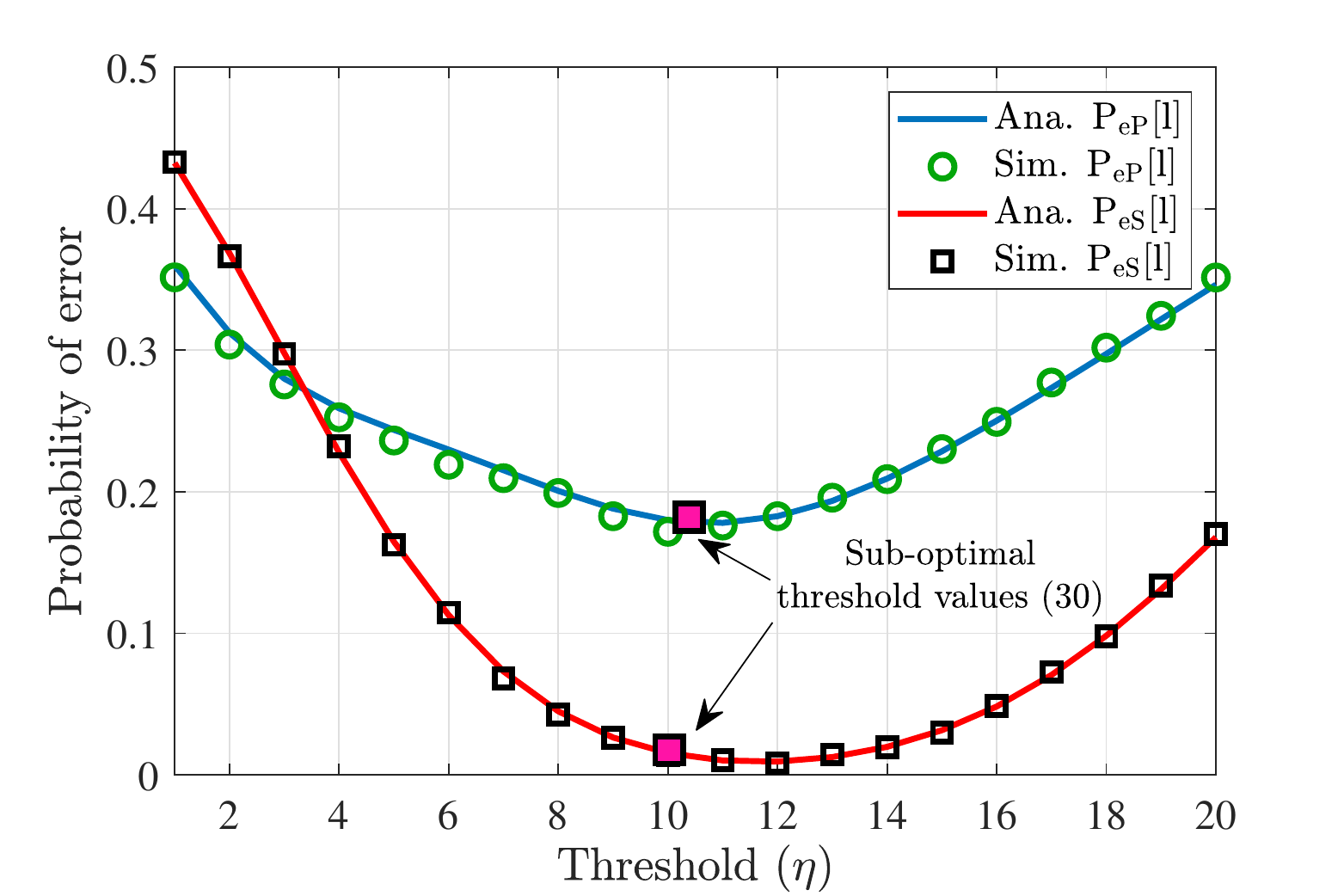}
	\caption{Variation of the probability of bit error with respect to threshold $ \eta $. Parameters: $D=100\mu m/s,\ a_\p=3\mu m, a_\s=5 \mu m, \ts=2 s, l=3, u_\m=5, u_\mathrm{L}=300, \mu=0.5s^{-1},\ \x_\p{=}[30, -10, 0],\ \x_\s{=}[10, 10, 0], 
		\ \y_\p=[30,10,0],\ \y_\s=[10, 10, 20] $.}
	\label{fig:p_eta}
\end{figure}
Fig. \ref{fig:p_eta} shows the variation of probability of bit error with the threshold for detection $ \eta $ at the FAR$_\pu  $ and the  FAR$_\su  $. With the increase in $ \eta $, the probability of bit error first reduces until it reaches the minimum point and then starts increasing. Therefore, there exists an optimal threshold for which the probability of bit error is minimum. In Fig. \ref{fig:p_eta}, the probability of bit error at the FAR$_\pu  $ is higher than that of FAR$_\su  $ since the size of FAR$_\pu  $ is smaller, and thereby the hitting probability is smaller compared to that of FAR$_\su  $. \par
 It is evident from Fig. \ref{fig:p_eta} that the detection threshold plays a vital role in determining the performance of the system and must be chosen correctly to increase system performance which we study next. 
\subsection{Detection threshold $ \eta[i;l] $}
 The probability of bit error derived in Theorem \ref{th2} varies significantly with the threshold of detection as seen in Fig. \ref{fig:p_eta}. The communication will be effective when the detection threshold is chosen such that the probability of bit error is minimum. Threshold for which the probability of bit error is minimum is termed as \textit{optimum threshold}. In this section, we derive the sub-optimum threshold for detection at the FAR$_\pu  $ and FAR$_\su  $.\par
   Now, for decoding at the FAR$ _i $, we can chose the  null hypothesis $ \mathcal{H}_{0i}[l] $ and alternative hypothesis $ \mathcal{H}_{1i}[l]$ as the event of sending bit $0 $ and $ 1 $ respectively by the TX$ _i $.
   The decision rule at the $ i $th receiver is
   \begin{align}
   	z_i[l]\overset{\mathcal{H}_{1i}}{\underset{\mathcal{H}_{0i}}{\gtrless}}\eta[i;l].\label{llrt2}
   \end{align} 
The calculation of $ \eta[i;l], \ i\in \{\pu, \su\} $ is complex since the received number of molecules are Binomial distributed. For tractability,  Binomial random variables $ z_{ i\noti}[l;k] \sim\mathcal{B}\left(b_i[k]u_i[k],h_{i\noti}[l-k]\right),\ 1\leq k\leq l, \ i\in \{\pu, \su\}, u_i[k]\in\{N,\us[r_\rsp;k]\} $ can be approximated to Poisson random variable $ z_{ i\noti}[l;k]\sim \mathcal{P}\left(b_i[k]u_i[k]h_{i\noti}[l-k]\right)$ \cite{Jamali2019a}.\par
 Using the Poisson approximation of $ z_{ i\noti}[l;k] $, the mean number of molecules absorbed at the FAR$ _i $ at the $ l $th time-slot when the transmitted bit is $ 0 $ and $ 1 $ are
	\begin{align}
	\lambda_{0i}[l]=N\qpro{1}{i}\sum_{k=1}^{l-1}u_i[k]h_{ii}[l-k]+\qpro{1}{\noti}\sum_{k=1}^{l}u_\noti[k]h_{\noti i}[l-k],
\end{align}
and
\begin{align}
	\lambda_{1i}[l]=u_i[l]h_{ii}[0]+\lambda_{0i},
\end{align}
respectively. For finding the sub-optimum decision threshold at receivers, the log-likelihood ratio test (LLRT) can be used,
\begin{align}
	\ln\left[\frac{\mathrm{P}\left(z_i[l]\mid \mathcal{H}_{1i}[l] \right)}{\mathrm{P}\left(z_i[l]\mid \mathcal{H}_{0i}[l] \right)}\right]\overset{\mathcal{H}_{1i}}{\underset{\mathcal{H}_{0i}}{\gtrless}}\ln\left[\frac{\qpro{0}{i}}{\qpro{1}{i}}\right],\label{llrt1}
\end{align}
where
\begin{align*}
	\mathrm{P}\left(z_i[l]\mid \mathcal{H}_{1i}[l] \right)=e^{-\lambda_{1i}[l]}\frac{\lambda_{1i}[l]^{z_i[l]}}{z_i[l]!},
\end{align*}
and
\begin{align*}
	\mathrm{P}\left(z_i[l]\mid \mathcal{H}_{0i}[l] \right)=e^{-\lambda_{0i}[l]}\frac{\lambda_{0i}[l]^{z_i[l]}}{z_i[l]!}.
\end{align*}

Now, solving \eqref{llrt1} and comparing with \eqref{llrt2} gives \cite{Chouhan2019}
 \begin{align}
 	\eta[i;l]=\frac{\ln\left(\qpro{0}{i}/\qpro{1}{i}\right)+h_{ii}[0]u_i[l]}{\ln\left(\lambda_{1i}[l]/\lambda_{0i}[l]\right)}.\label{opt_thr}
 \end{align}
The detection threshold value derived in \eqref{opt_thr} is the sub-optimal threshold for the FAR$_\pu  $ and FAR$_\su  $ at the $ l $th time-slot. It is evident from Fig. \ref{fig:p_eta} that, the sub-optimal values obtained in \eqref{opt_thr} is close to the optimal threshold values. The derived value of detection threshold is sub-optimal since it is calculated based on the average value instead of instantaneous values.
\subsection{Effect of Fixed and Variable Threshold on Probability of Bit Error}
\begin{figure}
	\centering
	\includegraphics[width=0.6\linewidth]{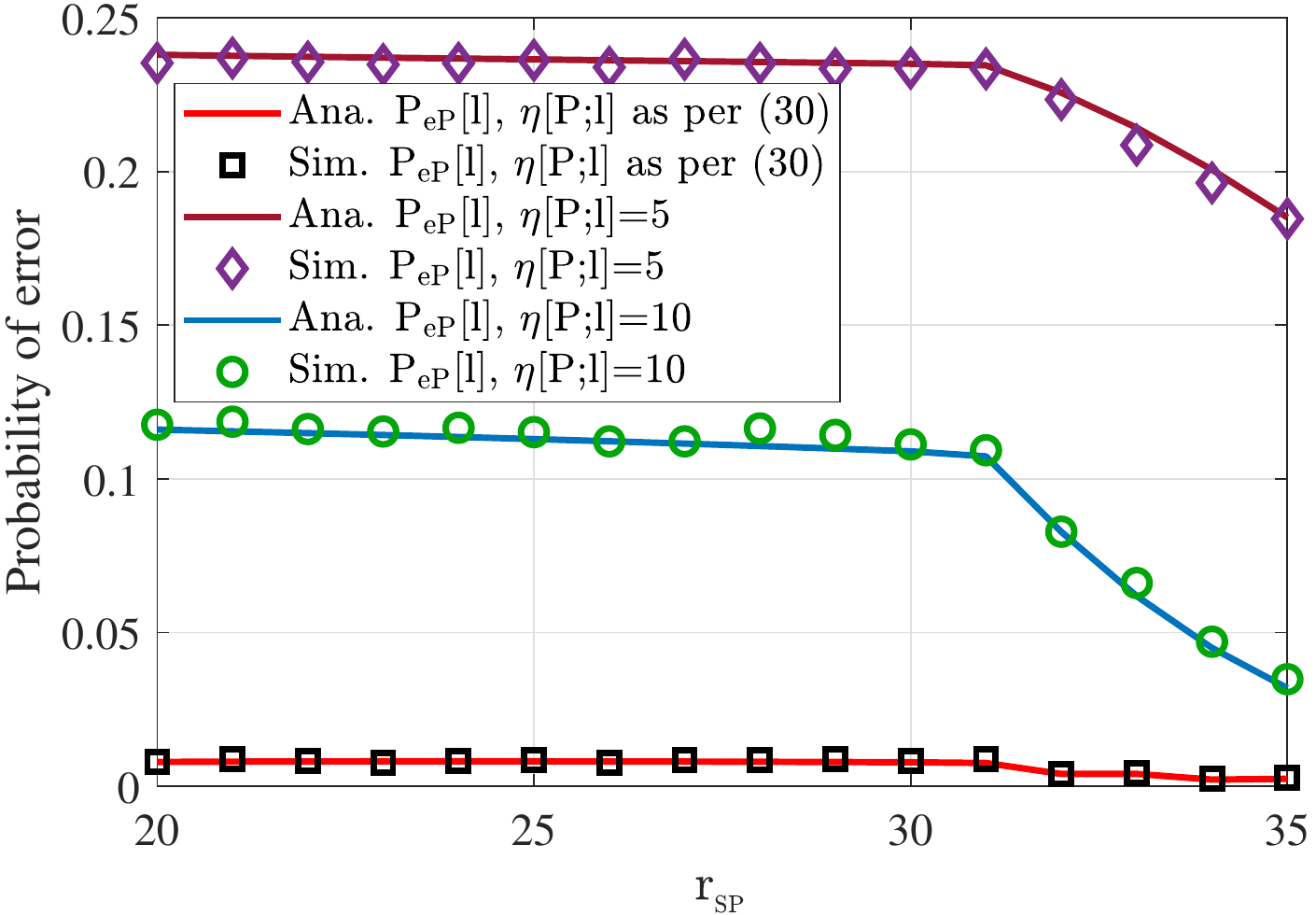}
	\caption{Variation of the probability of bit error with respect to $r_\rsp$ when $ \eta[i;l] $ is varying as in \eqref{opt_thr} or fixed.  Parameters: $D=100\mu m/s,\ a_\p=5 \mu m,\ a_\s=5 \mu m,\ \ts=5 s,\ l=3,\ u_\m=5,\ u_L=300,\ \x_\p=[30, -10, 0], 
		\ \y_\p=[30,10,0],\ \y_\s=[10, 10, 20] $. The initial location of $ \x_\s=[0, 10, 0]$ and it is varied in positive $ x-$ axis.}
	\label{fig:p_pu_deg}
\end{figure}
Fig. \ref{fig:p_pu_deg} shows the variation of probability of bit error at the FAR$_\pu  $ with respect to $ r_\rsp $ for fixed $ \eta $ and sub-optimal $ \eta[i;l] $ values. The probability of bit error is minimum when the $ \eta[i;l] $ values are optimally controlled. Fixed values of $ \eta $ result in a higher probability of bit error. When $ r_\rsp $ is high, the co-channel interference molecules reaching the FAR$_\pu  $ is less than the maximum allowable interference $ u_\m $. When $ r_\rsp $ reduces, co-channel interference increases and reaches $ u_\m $ eventually, thereby maintaining the probability of bit error constant afterward.
\subsection{Effect of $ r_\rsp $ on Probability of Bit Error}
\begin{figure}
	\centering
	\includegraphics[width=0.7\linewidth]{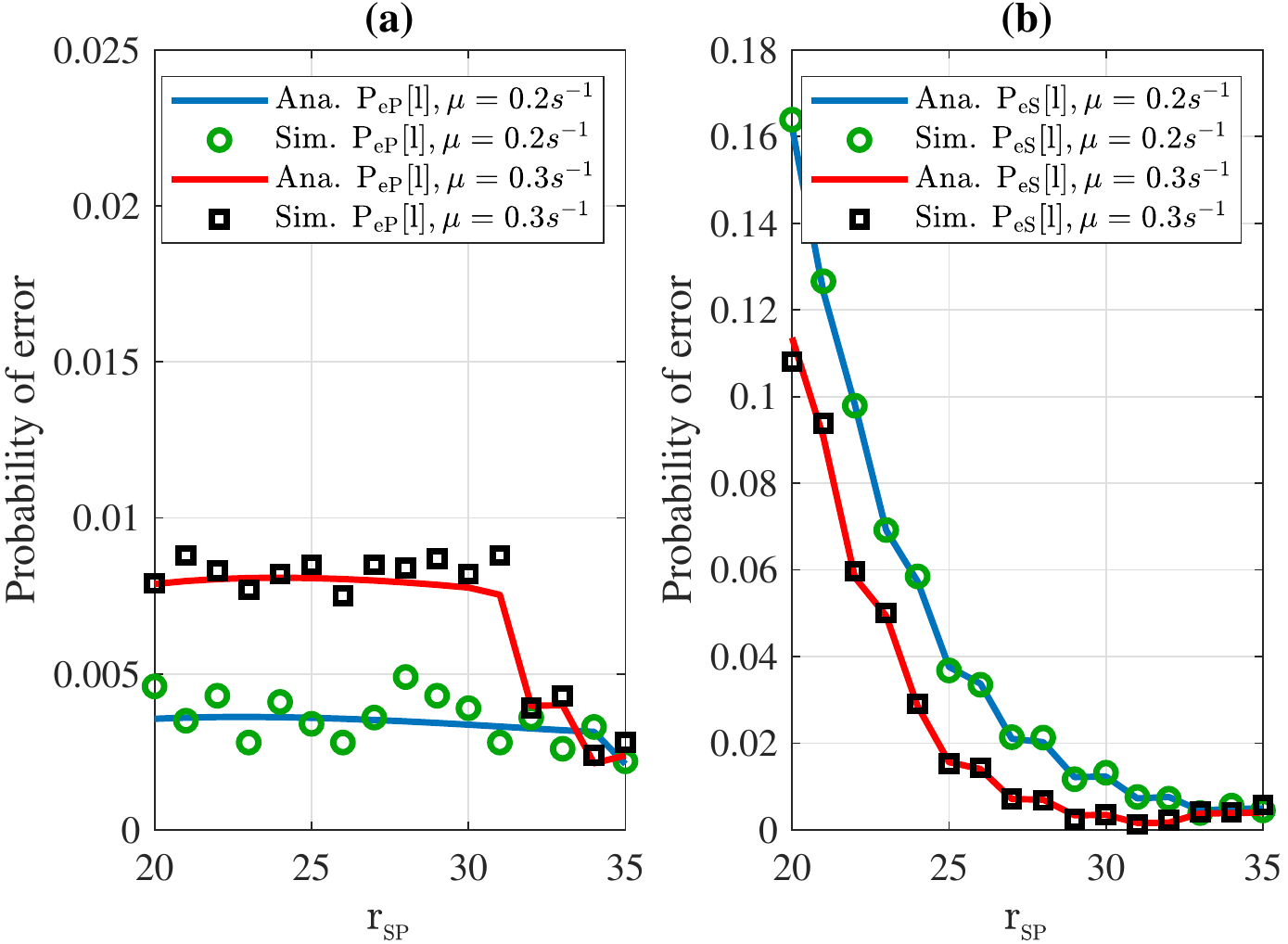}
	\caption{Variation of the probability of bit error with respect to $ r_\rsp $ for different degradation values. Parameters: $D=100\mu m/s,\ a_\p=5\mu m, a_\s=5 \mu m, \ts=5 s, l=3, u_\m=5, u_L=300,\ \x_\p=[30, -10, 0],
		\ \y_\p=[30,10,0],\ \y_\s=[10, 10, 20],\ \eta[i;l]=\text{as per \eqref{opt_thr}} $. The initial location of $ \x_\s=[0, 10, 0]$ and it is varied in positive $ x-$ axis}
	\label{fig:p_deg}
\end{figure}
Fig. \ref{fig:p_deg} shows the variation of probability of bit error with the distance between TX$_\su  $ and FAR$_\pu  $ ($ r_\rsp $) for two different values of molecular degradation rate constant. When $ r_\rsp $ decreases, the co-channel interference at the FAR$_\pu  $ increases and the probability of bit error rises. However, due to the controlling of the transmitted number of molecules by the TX$_\su  $, the co-channel interference is limited to $ u_\m $, and the probability of bit error remains unaffected. Nevertheless, due to this restriction, when $ r_\rsp $ reduces, $\us[r_\rsp;l]$ reduces, and the probability of bit error at FAR$_\su  $ increases. Therefore, for the cognitive pair to work with good performance, TX$_\su  $ should be far from FAR$_\pu  $.\par
When molecular degradation is faster, the primary and secondary link performance changes in different ways, as seen in Fig. \ref{fig:p_deg} (a) and (b). For the chosen parameters, TX$_\pu  $ to FAR$_\pu  $ distance and TX$_\su  $ to FAR$_\su  $ distance is less than  TX$_\pu  $ to FAR$_\su  $ distance and TX$_\su  $ to FAR$_\pu  $ distance. Therefore co-channel interference molecules are more prone to degradation. Secondary link performance improves with degradation because the number of interfering molecules to FAR$_\pu  $ degrades more. Due to this, TX$_\su  $ can send more molecules to FAR$_\su  $ without crossing the co-interference beyond the desired limit of $ u_\m $. For the primary link, the co-channel interference does not change with $ \mu $ when FAR$_\pu  $ and TX$_\su  $ are nearby, but at the same time, the desired molecules emitted by TX$_\pu  $ to FAR$_\pu  $ degrades in a small amount. This results in the slight diminishing of bit error performance when $ \mu $ is increased. \subsection{Effect of Controlled Transmission on Probability of Bit Error}  
\begin{figure}
	\centering
	\includegraphics[width=0.7\linewidth]{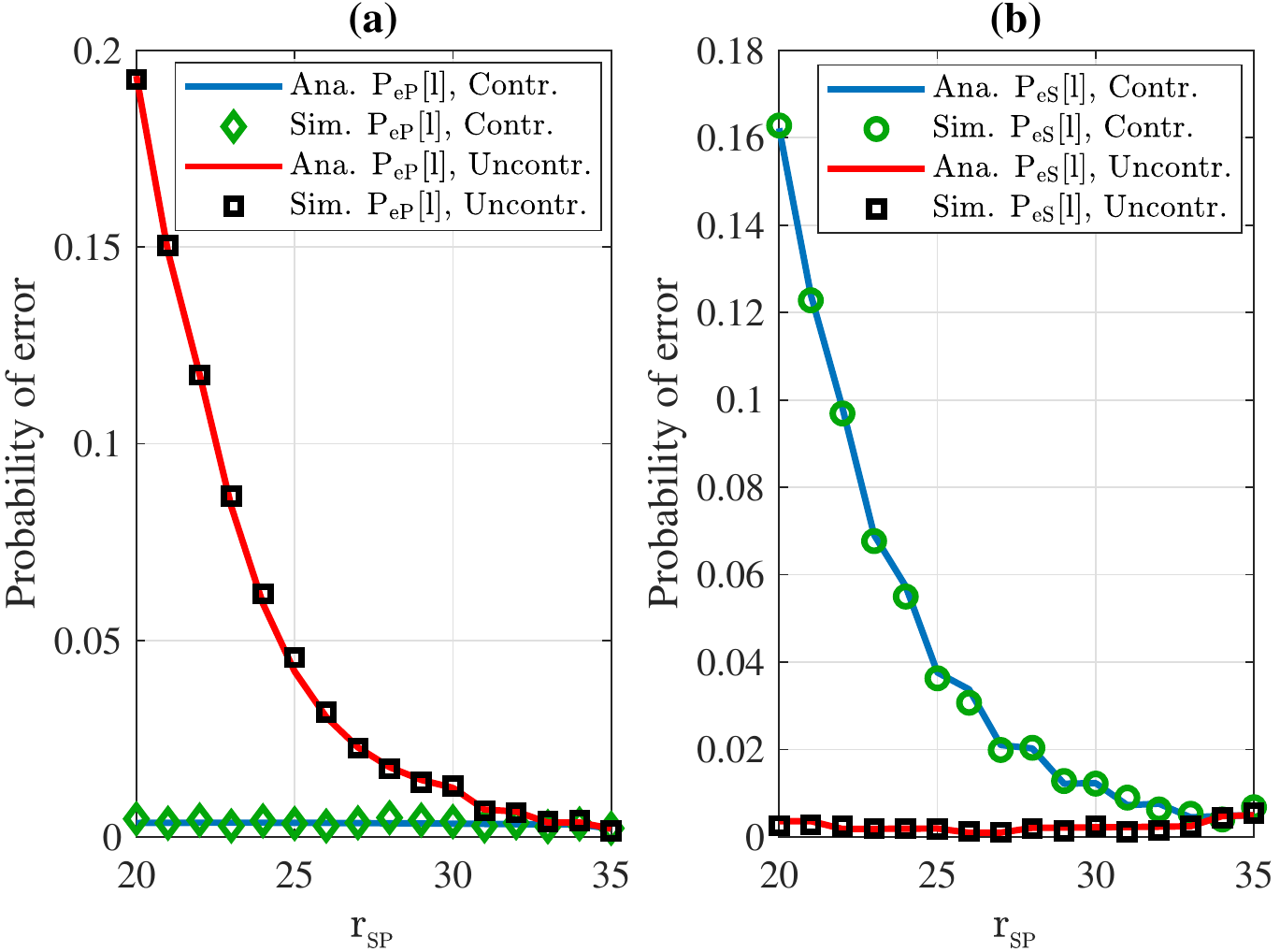}
	\caption{Variation of the probability of bit error with respect to $r_\rsp$ when $ \us[r_\rsp;l] $ is controlled and uncontrolled.  Parameters: $D=100\mu m/s,\ a_\p=5 \mu m,\ a_\s=5 \mu m, \ts=5 s, l=3, u_\m=5, u_L=300,\ \x_\p=[30, -10, 0],	\ \y_\p=[30,10,0],\ \y_\s=[10, 10, 20],\ \eta[i;l]=\text{as per \eqref{opt_thr}}$.  The initial location of $ \x_\s=[0, 10, 0]$ and it is varied in positive $ x-$ axis.}
	\label{fig:p_fix}
\end{figure}
Fig. \ref{fig:p_fix} shows the effect of the control of molecules emitted by the TX$_\su  $ on the bit error probability. In the controlled emission of the molecules from the TX$_\su  $, molecule emission is controlled according to $ r_\rsp $  as in \eqref{ustran} for bit 1 transmission. However, for uncontrolled emission the  TX$_\su  $ emits a fixed number (here $\us[r_\rsp;l]=300$ molecules) of molecules for bit 1 transmission. Fig. \ref{fig:p_fix} (a) and (b) validates that the controlling of molecules emitted by the TX$_\su  $ based on $ r_\rsp $ helps to maintains good performance to the high priority primary link. In contrast, the lower priority cognitive link performance deteriorates when $ r_\rsp $ reduces.
\section{Conclusions}
In this work, we consider a cognitive molecular communication system with an underlay strategy for the co-existence of two communication links with different priorities. Each link has a point transmitter and spherical fully-absorbing receiver communicating with the same type of molecules. The transmitted number of molecules by the low priority secondary transmitter is controlled to limit its co-channel interference to the high priority primary link.
 We first derive an approximate equation for the information molecule hitting probability on each of the spherical fully-absorbing receivers with different sizes. We include the impact of molecular degradation in our analysis. The derived equation is used to develop an analytic framework for the cognitive molecular communication. The underlay strategy of cognitive molecular communication is studied, and several essential insights have been presented in this work.

\appendices
\section{Proof of Remark \ref{r4}}\label{steady}

Recall that, the average number of interfering molecules should not exceed an acceptable threshold $u_\m$. At the steady state,
\begin{align}
	\expect{}{C_\rsp[\infty]} = \expect{}{\sum_{k=1}^{\infty} z_\rsp[l;k]} \leq u_\m. \label{AN1}
\end{align}
The $\expect{}{C_\rsp[\infty]}$ in above expression can be further solved as 
\begin{align}
	\expect{}{C_\rsp[\infty]} &= \sum_{k=1}^{\infty} \expect{}{z_\rsp[l;k]}\nonumber\\
	&=\qpro{1}{\su} \us[r_\rsp;\infty]\sum_{k=0}^{\infty} h_\rsp[k]\nonumber\\
	&=\qpro{1}{\su} \us[r_\rsp;\infty]\ptwo{\s\p}{}{\infty,\mu}. \label{AN2}
\end{align}
Comparing \eqref{AN1} and \eqref{AN2} gives \eqref{eqsteady}.

\section{Proof of Theorem \ref{t1}}\label{app:A}
The probability that the IM emitted by the point source located at $ \x_m $ reaches FAR$ _\noti $ in the interval $ [\tau,\ \tau+\dd\tau] $ is
$ \left[\frac{\partial \ptwo{m\noti}{}{\tau}}{\partial\tau}\right]\dd \tau $.  The probability that this IM hits the FAR$ _i $ in the remaining time $ t-\tau $ is $  \pone{i}{t-\tau,a,R_{m\noti i}} $, where $ R_{m\noti i} $ is the distance between the nearest point on the surface of FAR$ _\noti $ from the transmitter (located at $ \x_m $) and the center of FAR$ _i $ ($ E_\noti $ as seen in Fig.  \ref{fig:hp}).  Here the initial location of IM generation after the hit at FAR$ _\noti $ is approximated as the nearest point on the surface of FAR$ _\noti$ ($E_\noti$) from the transmitter. The probability of an IM that is supposed to hit the FAR$ _i $ within time $ t $ but is hitting FAR$ _\noti $ instead is 
\begin{align}
\pone{}{t,a_i,r_{mi}}{-} \ptwo{mi}{}{t}{=}\int_{0}^{t}\frac{\partial \ptwo{m\noti}{}{\tau}}{\partial\tau}\pone{}{t-\tau,a_i,R_{m\noti i}} \dd\tau.\label{ae1}
\end{align}
Similarly, the probability of an IM that is supposed to hit the FAR$ _j $ with in time $ t $ but is hitting the FAR$ _i $ is
\begin{align}
\pone{}{t,a_{\noti},r_{m\noti}}{-} \ptwo{m\noti}{}{t}=\int_{0}^{t}\frac{\partial \ptwo{mi}{}{\tau}}{\partial\tau}\pone{}{t-\tau,a_{\noti},R_{mi\noti}} \dd\tau.\label{ae2}
\end{align}
Now, taking the Laplace transform of \eqref{ae1} and \eqref{ae2} gives
\begin{align}
\lpone{}{s,a_i,r_{mi}}{-} \lptwo{mi}{}{s}{=}s\lptwo{m\noti}{}{s}\lpone{}{s,a_i,R_{m\noti i}}\label{ae3}
\end{align} and 
\begin{align}
\lpone{}{s,a_{\noti},r_{m\noti}}{-} \lptwo{m\noti}{}{s}{=}s\lptwo{mi}{}{s}\lpone{}{s,a_{\noti},R_{mi\noti}}\label{ae4}
\end{align}
Note that \cite{Yilmaz2014},
\begin{align}
	\pone{}{t,x,y}=\frac{x}{y}\erfc\left(\frac{y-x}{\sqrt{4Dt}}\right),\label{gen_eq1}
\end{align}
and its Laplace transform is
\begin{align}
	\lpone{}{s,x,y}=\frac{x}{y}\frac{\exp\left(-\left(y-x\right)\sqrt{\frac{s}{D}}\right)}{s},\label{gen_eq}
\end{align} where $ \ x\in\{a_i,a_{\noti}\},\ y\in\{r_{mi},r_{m\noti},R_{mi\noti},R_{m\noti i}\} $.
Solving \eqref{ae3} and \eqref{ae4} gives 
\begin{align}
	\lptwo{mi}{}{s}=&\frac{\lpone{}{s,a_i,r_{mi}}-s\lpone{}{s,a_{\noti},r_{m\noti}}\lpone{}{s,a_i,R_{m\noti i}}}{1-s^2\lpone{}{s,a_i,R_{m\noti i}}\lpone{}{s,a_{\noti},R_{mi\noti}}}\nonumber\\
    =&\sum_{n=0}^{\infty}\left(\lpone{}{s,a_i,r_{mi}}-s\lpone{}{s,a_{\noti},r_{m\noti}}\lpone{}{s,a_i,R_{m\noti i}}\right)\nonumber\\
    &\spc\spc\times s^{2n}\lpone{}{s,a_i,R_{m\noti i}}^n\lpone{}{s,a_{\noti},R_{mi\noti}}^n\label{aeq36}
\end{align}
Now, the direct substitution of the values of Laplace transforms in \eqref{aeq36} using \eqref{gen_eq} gives,
\begin{align}
	&\lptwo{mi}{}{s}=
	\sum_{n=0}^{\infty}\frac{a_i}{r_{mi}}\left(\frac{a_i}{R_{m\noti i}}\right)^{n}\left(\frac{a_{\noti}}{R_{mi\noti}}\right)^{n}\frac1s\exp\left(-( r_{mi}-a_i\right.
	\left.+n(R_{m\noti i}{-}a_i)+n(R_{mi\noti}{-}a_{\noti}))\sqrt{\frac{s}{D}}\right)\nonumber\\
	&\spc-\sum_{n=0}^{\infty}\frac{a_{\noti}}{r_{m\noti}}\left(\frac{a_i}{R_{m\noti i}}\right)^{n+1}\left(\frac{a_{\noti}}{R_{mi\noti}}\right)^{n}\frac1s\exp\left(-( r_{m\noti}-a_{\noti}\right.
	\left.+(n+1)(R_{m\noti i}{-}a_i)+n(R_{mi\noti}{-}a_{\noti}))\sqrt{\frac{s}{D}}\right)\label{ae5}
\end{align}
Taking the inverse Laplace transform of \eqref{ae5} gives Theorem \ref{t1}.

\section{Proof of Theorem \ref{t2}}\label{app:deg}
The fraction of non-degraded information molecule reaching the FAR$_i$ in the presence of the other FAR within time $ t $, due to the emission of IMs from the transmitter at $\x_{m}$ is given by
\begin{align}
\ptwo{mi}{}{t,\mu}
=&\int_{0}^{t}\hr{mi}{\tau} \exp\left(-\mu \tau\right)\dd \tau\nonumber\\
=&\int_{0}^{t}\sum_{n=0}^{\infty}\left(\frac{a_ia_{\noti}}{R_{mi\noti}R_{m\noti i}}\right)^{n}
\times\left[\frac{a_i}{r_{mi}}
\frac{\Phi_{mi}(n)}{\sqrt{4\pi D \tau^3}}\right.\exp\left( -\frac{\Phi_{mi}(n)^2}{4D\tau}-\mu \tau\right)\nonumber\\
&\spc\spc-\left.\frac{a_ia_{\noti}}{r_{m\noti}R_{m\noti i}}
\frac{\Psi_{mi}(n)}{\sqrt{4\pi D \tau^3}}\exp\left( -\frac{\Psi_{mi}(n)^2}{4D\tau}-\mu \tau\right)\right]\dd \tau\label{be1}
\end{align}
Note that,
\begin{align}
	\int_{0}^{t}\frac{y}{x^{3/2}}\exp\left(-\frac{y^2}{x}-zx\right)\dd x&=\frac{\sqrt{\pi}}{2}\left[\erfc\left(\frac{y}{\sqrt{t}}{+}\sqrt{z t}\right)\exp\left(2y\sqrt{z}\right)
	\right.\nonumber\\
	&\left.\spc+\erfc\left(\frac{y}{\sqrt{t}}{-}\sqrt{z t}\right)\exp\left({-}2y\sqrt{z}\right)\right].\label{be2}
\end{align}
Now, applying \eqref{be2} on \eqref{be1} gives Theorem \ref{t2}.
\section{Proof of Theorem \ref{th2}}\label{app:B}
The probability of incorrect decoding of bit 1 at the FAR$_\pu  $ is given by
\begin{align}
\pe{\mathrm{P},1}[l]&=\pro\left(z_\p[l]\leq \eta[\pu;l]\mid \symp[l]=1\right)\label{pea1}. 
\end{align}
Now, using the identity in \cite[eq. 1.272]{Johnson2005}, \eqref{pea1} can be written as
\begin{align}
\pe{\mathrm{P},1}[l]&=\sum_{n=0}^{\eta[\pu;l]-1}\frac{\pgfa{(n)}{0}}{n!}\label{pea}, 
\end{align}
where $\pgfa{}{v}=\expect{z_\p[l]\mid \symp[l]=1}{v^{z_\p[l]}}$. Here, $\pgfa{}{v}$  is the probability generating function (PGF) \cite{Beaumont2005} of $z_\p[l]$ and $\pgfa{(n)}{0}$ represents the $n$th derivative of $ \pgfa{}{v} $ with $v=0$.
\subsubsection{PGF Calculation}
 The PGF of $z_\p[l]$ when $ b_\p[l]=1 $ is 
\begin{align}
\pgfa{}{v}&=\expect{z_\p[l]\mid \symp[l]=1}{v^{z_\p[l]}}\nonumber\\
&=\expect{}{v^{z_\rpp[l;l]+\sum_{k=1}^{l-1}z_\rpp[l;k]+\sum_{k=1}^{l}z_\rsp[l;k]}}\nonumber\\
&=\expect{}{v^{z_\rpp[l;l]}}\expect{}{v^{\sum_{k=1}^{l-1}z_\rpp[l;k]}}\expect{}{v^{\sum_{k=1}^{l}z_\rsp[l;k]}}\nonumber\\
&\stackrel{\text{(a)}}{=}\expect{}{v^{z_\rpp[l;l]}}\prod_{k=1}^{l-1}\expect{}{v^{z_\rpp[l;k]}}\prod_{k=1}^{l}\expect{}{v^{z_\rsp[l;k]}}.\label{e33}
\end{align}
Here $ (a) $ is due to the independence of $ z_\rpp[l;k] $ and $ z_\rsp[l;k] $ across $ l $ and $ k $ \cite{Johnson2005}. Note that, the PGF of random variable $ X $ with Binomial distribution $ \mathcal{B} (n,p)$ is $ \mathcal{G}(z)=\expect{}{z^X}=\left(1-p+pz\right)^n $. Hence, the product terms in \eqref{e33} can be obtained as
\begin{align}
\expect{}{v^{z_\rpp[l;l]}}&=\left[1-h_\rpp[0]+h_\rpp[0]v\right]^{N}\nonumber\\
&=\left[1+h_\rpp[0]\left(v-1\right)\right]^{N},\label{e34}
\end{align}
\begin{align}
\expect{}{v^{z_\rpp[l;k]}}&=\pro\left[\symp[k]=0\right]\expect{}{v^{z_\rpp[l;k]}\mid\symp[k]=0}\nonumber\\
&+\pro\left[\symp[k]=1\right]\expect{}{v^{z_\rpp[l;k]}\mid\symp[k]=1} \nonumber\\  &=\left[\qpro{0}{\pu}+\qpro{1}{\pu}\expect{}{v^{z_\rpp[l;k]}}\right]\nonumber\\
&=\left[\qpro{0}{\pu}+\qpro{1}{\pu}\left(1+h_\rpp[l-k](v-1)\right)^{N}\right],\label{e35}
\end{align}
and
\begin{align}
\expect{}{v^{z_\rsp[l;k]}}=&\pro\left[\syms[k]=0\right]\expect{}{v^{z_\rsp[l;k]}\mid\symp[k]=0}\nonumber\\
&+\pro\left[\syms[k]=1\right]\expect{}{v^{z_\rsp[l;k]}\mid\symp[k]=1} \nonumber\\ =&\left[\qpro{0}{\su}+\qpro{1}{\su}\left(1+h_\rsp[l-k](v-1)\right)^{\us[r_\rsp;k]}\right]\label{e36}.
\end{align}

Substituting \eqref{e34}, \eqref{e35} and \eqref{e36}, in \eqref{e33} gives
\begin{align}
\pgfa{}{v}=\left[1+h_\rpp[0]\left(v-1\right)\right]^{N}
&\times \prod_{k=1}^{l-1}\left[\qpro{0}{\pu}+\qpro{1}{\pu}\left(1+h_\rpp[k](v-1)\right)^{N}\right]\nonumber\\
&\times\prod_{\substack{k=0,\\ \us[r_\rsp;l-k]\neq 0}}^{l-1}\left[\qpro{0}{\su}+\qpro{1}{\su}\left(1+h_\rsp[k](v-1)\right)^{\us[r_\rsp;l-k]}\right]\label{e37}
\end{align} 

\subsubsection{$n$th derivative of PGF}
From \eqref{e37}, let $ \theta_k(v) $ be defined as
\begin{align}
&\theta_k(v)=
\begin{cases}
\left(1+h_\rpp[0]\left(v-1\right)\right)^{N},&\rem\rem\text{if } k=0,\\
\qpro{0}{\pu}+\qpro{1}{\pu}\left(1+h_\rpp[k](v-1)\right)^{N}&\rem\rem\text{if } 1\leq k\leq l-1,\\
\qpro{0}{\su}+\qpro{1}{\su}\left(1+h_\rsp[k-l](v-1)\right)^{\us[r_\rsp;2l-k]},&\rem\rem\text{if } l\leq k\leq 2l-1.
\end{cases}\label{ai}
\end{align}
Note that, $ \pgfa{}{v}=\prod_{k=0}^{2l-1}\theta_k(v) $. The $n$th derivative of the PGF derived in \eqref{e37} can be obtained by using General Leibniz rule \cite{Sohrab2014}, that is,
\begin{align}
\pgfa{(n)}{v}&=\frac{\diff^n}{\diff v^n}\left(\prod_{k=0}^{2l-1}\theta_k(v)\right)\nonumber\\
&=\sum_{\substack{n_0+n_1+\cdots+n_{2l-1}=n\\ n_k<u_\s[r_\rsp;2l-k],\ \forall l\leq k\leq 2l-1}}{n\choose n_0,n_1,\ldots,n_{2l-1}}\prod_{k=0}^{2l-1}\theta_k^{(n_k)}(v)\label{An},
\end{align}
where $ \theta_k^{(n_k)}(v) $ is the $ n_k $th derivative of  $ \theta_k(v) $, which is defined as
\begin{align}
&\theta_k^{(n_k)}(v)=
\begin{cases} 
\left(1+h_\rpp[0]\left(v-1\right)\right)^{N-n_k}&\text{if }k=0,\\ 
\times h_\rpp[0]^{n_k}\frac{N!}{(N-n_k)!}
\\
\qpro{1}{\pu}\left(1+h_\rpp[k]\left(v-1\right)\right)^{N-n_k}&\text{if }1\leq k\leq l-1,\\ 
\times  h_\rpp[k]^{n_k}\frac{N!}{(N-n_k)!} \\ 
\qpro{1}{\su}\left(1+h_\rsp[k-l]\left(v-1\right)\right)^{\us[r_\rsp;2l-k]-n_k} &\text{if } l\leq k\leq 2l-1,\\
 \times  h_\rsp[k-l]^{n_k} \frac{\us[r_\rsp;2l-k]!}{\left(\us[r_\rsp;2l-k]-n_k\right)!}& k<\us[r_\rsp;2l-k] .
\end{cases}\label{an}
\end{align}
Substituting \eqref{an} in \eqref{An} with $ v=0 $ in \eqref{pea} gives \eqref{Ap}.\par
Similarly, the error probability of bit 0 at the FAR$_\pu  $ is is given as
\begin{align}
	\pe{\mathrm{P},0}[l]&=1-\pro\left(z_\p[l]\leq \eta[\pu;l]\mid \symp[l]=0\right)\nonumber\\
	&=1-\sum_{n=0}^{\eta[\pu;l]-1}\frac{\pgfb{(n)}{0}}{n!}\label{peb},
\end{align}
where $\pgfb{}{v}=\expect{z_\p[l]\mid \symp[l]=0}{v^{z_\p[l]}}$ is the PGF of $z_\p[l]$ and $\pgfb{(n)}{0}$ represents the $n$th derivative of $\pgfb{}{v}$ with $v=0$.
 The PGF of $z_\p[l]$ when $ b_\p[l]=0 $ is 
\begin{align}
\pgfb{}{v}&=\expect{z_\p[l]\mid \symp[l]=0}{v^{z_\p[l]}}\nonumber\\
&=\prod_{k=1}^{l-1}\expect{}{v^{z_\rpp[l;k]}}\prod_{k=1}^{l}\expect{}{v^{z_\rsp[l;k]}}.\label{bz}
\end{align}
Solving \eqref{bz} further and using General Leibniz rule as in the derivation of \eqref{An} gives
\begin{align}
\pgfb{(n)}{v}&=\rem\sum_{\substack{n_1+n_2+\cdots+n_{2l-1}=n\\ n_k<u_\s[r_\rsp;2l-k],\ \forall l\leq k\leq 2l-1}}{n\choose n_1,n_2,\ldots,n_{2l-1}}\prod_{k=1}^{2l-1}\rho_k^{(n_k)}(v)\label{Bn},
\end{align}
where
\begin{align}
	&\rho_k^{(n_k)}(v)=
	\begin{cases} 
		\qpro{1}{\pu}\left[1+h_\rpp[k]\left(v-1\right)\right]^{N-n_k}& \text{if }1\leq k\leq l-1, \\ \times  h_\rpp[k]^{n_k}\frac{N!}{(N-n_k)!}\\ 
		\qpro{1}{\su}\left[1+h_\rsp[k-l]\left(v-1\right)\right]^{\us[r_\rsp;2l-k]-n_k}& \text{if }l\leq k\leq 2l-1,\\
		\times  h_\rsp[k-l]^{n_k}\frac{\us[r_\rsp;2l-k]}{\left(\us[r_\rsp;2l-k]-n_k\right)!}& k<\us[r_\rsp;2l-k] . 
	\end{cases}\rem.\label{bn}
\end{align}
Substituting \eqref{bn} in \eqref{Bn} with $ v=0 $ in \eqref{peb} gives \eqref{Bp}.

	\bibliographystyle{IEEEtran}
	\bibliography{Cognitive_ref}

\end{document}

%% file: define.tex
\ifCLASSINFOpdf
\usepackage[pdftex]{graphicx}
\graphicspath{{figs/}}
\DeclareGraphicsExtensions{.pdf}
\else
\fi

\usepackage{mathtools}
\usepackage{graphicx}
\usepackage{amsthm}
\usepackage{cite}
\usepackage{color}
\usepackage{bbm}
\usepackage{amsmath,amssymb,amsfonts}
\usepackage{subcaption} 
\DeclarePairedDelimiter\floor{\lfloor}{\rfloor}
\newcommand{\qpro}[2]{q_{_{#1#2}}}

\newcommand{\ie}{\textit {i.e., }}
\newcommand{\x}{\pmb{x}}
\newcommand{\p}{{{}_{\mathrm{P}}}}
\newcommand{\pu}{\mathrm{P}}
\newcommand{\su}{\mathrm{S}}
\newcommand{\m}{\mathrm{M}}

\newcommand{\s}{{{}_{\mathrm{S}}}}
\newcommand{\rpp}{{{}_{\mathrm{PP}}}}
\newcommand{\rss}{{{}_{\mathrm{SS}}}}
\newcommand{\rps}{{{}_{\mathrm{PS}}}}
\newcommand{\rsp}{{{}_{\mathrm{SP}}}}
\newcommand{\y}{\pmb{y}}
\newcommand{\rem}{\!\!\!\!}
\newcommand{\spc}{\ \ \ \ }
\newcommand{\h}[2]{h_{#1 #2}}

\newcommand{\pgfa}[2]{\mathrm{A}^{#1}\left(#2\right)}
\newcommand{\pgfb}[2]{\mathrm{B}^{#1}\left(#2\right)}

\newcommand{\up}{u_{\p}}
\newcommand{\us}{u_{\s}}

\newcommand{\symp}{b_\p}
\newcommand{\symi}{b_{i}}
\newcommand{\esymi}{\hat{b}_{i}}

\newcommand{\syms}{b_\s}
\newcommand{\ts}{T_{\mathrm{b}}}
\newcommand{\norm}[1]{\left\lVert #1\right \rVert}
\newcommand{\diff}{\mathrm{d}}

\newcommand{\erfc}{\mathrm{erfc}}

\newcommand{\expect}[2]{\mathbb{E}_{#1}\left[#2\right]}
\newcommand{\pro}{\mathbb{P}}
\newcommand{\pe}[1]{\mathrm{P}_{\mathrm{e}#1}}
\newcommand{\rthree}{\mathbb{R}^3}

\newcommand{\ptwo}[3]{\mathrm{p}_{#1}^{#2}\left({#3} \right)}
\newcommand{\pone}[2]{\overline{\mathrm{p}}_{}\left({#2} \right)}
\newcommand{\lptwo}[3]{\mathcal{P}_{#1}^{#2}\left({#3} \right)}
\newcommand{\lpone}[2]{\overline{\mathcal{P}}_{}\left({#2} \right)}
\newcommand{\hr}[2]{\kappa_{#1}\left({#2} \right)}

\newcommand{\dd}{\mathrm{d}}

\newcommand{\noti}{{\bar{i}}}
\newtheorem{theorem}{Theorem}

\newtheorem{coro}{Corollary}[theorem]
\newtheorem{remark}{Remark}